\documentclass[aps,prd,onecolumn,nofootinbib]{revtex4-2}

\usepackage{amsmath,amssymb,graphicx,hyperref,bm,xcolor}

\usepackage[a4paper,margin=1in]{geometry}

\usepackage{verbatim}
\usepackage{enumitem}
\usepackage{booktabs}
\usepackage{url}

\usepackage{kotex}

\newcommand{\wCDM}{w\textrm{CDM}}
\newcommand{\Omo}{\Omega_{\rm{m}0}}

\begin{document}

\title{Information-Geometric Perspective on the Hubble Tension: Eigenmode Rotation and Curvature Suppression in $w$CDM}

\author{Seokcheon Lee \email{skylee@skku.edu}}
\affiliation{Department of Physics, Institute of Basic Science, Sungkyunkwan University, Suwon 16419, Republic of Korea}

\title{Information-Geometric Perspective on the Hubble Tension: Eigenmode Rotation and Curvature Suppression in $w$CDM}

\begin{abstract}
The Hubble tension is determined not only by the relative shifts between early- and late-time measurements but also by the stiffness of the constraints along the relevant parameter combinations. We investigate this structure in the $w$CDM model by decomposing the discrepancy into a parameter displacement and a directional Fisher curvature in parameter space. Within the Gaussian approximation, the quadratic tension along a given direction factorizes into the squared parameter shift and the combined curvature contributed by the datasets. Applying this framework to Planck, DESI DR2, and SH0ES, we demonstrate that extending $\Lambda$CDM to $w$CDM primarily reshapes the Fisher geometry of the Planck constraint. Treating the dark energy equation of state $w$ as a free parameter suppresses the leading eigenvalue and rotates the dominant acoustic-scale eigenmode, thereby enlarging the existing degeneracy direction. Furthermore, our analysis of recent high-precision measurements such as DESI DR2 reveals that the increased statistical stiffness acts as a geometric wall, effectively blocking potential tension resolutions within standard model extensions. Consequently, the resulting change in the inferred $H_0$ tension is driven by a geometric reconfiguration of the constraint structure rather than the emergence of new independent information. This decomposition clarifies how parameter extensions modify cosmological constraints and provides a physically transparent way to interpret model-dependent variations in the level of the Hubble tension.
\end{abstract}

\maketitle

\tableofcontents

\section{Introduction}
\label{sec:introduction}
Determinations of the Hubble parameter have long been characterized by debates over calibration, methodology, and systematic uncertainty, with significant discrepancies noted even in pre-CMB measurements~\cite{Sandage:1975zz,deVaucouleurs:1982rrt}. The Hubble Space Telescope Key Project marked a major milestone in consolidating distance-ladder measurements, yet the precise value of $H_0$ remained sensitive to calibration choices and systematic effects~\cite{HST:2000azd,Sandage:2006cv}. Over the past decade, measurements of the present-day expansion rate have revealed a persistent discrepancy between early- and late-time probes~\cite{Verde:2013wza,Efstathiou:2013via,Bernal:2016gxb,Knox:2019rjx}. Cosmic microwave background (CMB) observations, most notably from \textit{Planck}, infer $H_0$ within the framework of $\Lambda$CDM at a value significantly lower than those obtained from local distance-ladder measurements such as SH0ES~\cite{Planck:2015fie,Planck:2018vyg,Riess:2011yx,Riess:2016jrr,Riess:2019cxk,Freedman:2019jwv,Riess:2020fzl,Riess:2021jrx}. Recent large-scale structure measurements, including DESI DR2, have further sharpened constraints on late-time geometry and stimulated renewed discussions on model extensions~\cite{DESI:2025zgx}.

This discrepancy, commonly referred to as the Hubble tension, is often quantified in terms of differences between best-fit parameters or shifts in posterior means expressed in units of the combined variance~\cite{Verde:2013wza,Efstathiou:2013via,Bernal:2016gxb,Knox:2019rjx}. A variety of statistical diagnostics have been formulated to assess dataset concordance, including parameter-difference statistics and index-of-inconsistency measures~\cite{Lin:2017ikq,Raveri:2018wln}, Bayesian evidence ratios for model comparison~\cite{Trotta:2008qt}, information-theoretic metrics based on relative entropy~\cite{Seehars:2014ora}, and suspiciousness-based measures of dataset tension~\cite{Handley:2019wlz}. While these approaches provide scalar measures of disagreement, they implicitly conflate two distinct ingredients: the parameter displacement and the directional stiffness (likelihood curvature) with which each dataset constrains the corresponding parameter combinations~\cite{Raveri:2018wln,Handley:2019wlz}. Therefore, the statistical significance of a shift depends not only on its magnitude, but also on the curvature of the likelihood along the direction of that displacement~\cite{Verde:2013wza,Lin:2017ikq}. Recent Bayesian reanalyses reinforce this point by demonstrating that scalar diagnostics often penalize model extensions through an Ockham factor, which reflects the inflation of parameter volume without a commensurate gain in the goodness-of-fit~\cite{Ormondroyd:2025phk,Ong:2025utx,Ong:2026tta}. Our information-geometric framework provides the physical underpinning for this statistical penalty by identifying the specific redistribution and stiffening of Fisher curvature under such extensions.

More generally, cosmological observables do not constrain individual parameters independently of one another.  Instead, they probe specific combinations of parameters through integral relations, projection effects, and geometric degeneracies~\cite{Hu:1995en,Tegmark:1996bz,Hu:1997hp,Efstathiou:1998xx,Eisenstein:1998hr,Albrecht:2006um,Raveri:2015maa,Addison:2015wyg,DiValentino:2021izs}. Thus, apparent discrepancies between datasets may reflect differences in how information about a given parameter direction is encoded in the likelihood geometry, rather than purely differences in central values~\cite{Raveri:2018wln,Handley:2019wlz,Trotta:2008qt}. 

In CMB analyses, $H_0$ is not a directly observed quantity but is inferred as a derived parameter within an assumed background cosmological model; it is therefore subject to the geometric degeneracy associated with the acoustic angular scale~\cite{Hu:1995en,Efstathiou:1998xx,Efstathiou:2013via,Hu:1997hp}. The CMB primarily constrains combinations such as $\theta_\ast \sim r_s/D_A$ rather than $H_0$ in isolation. Consequently, variations in $H_0$ can be partially compensated by correlated shifts in other background parameters, producing an extended degeneracy direction in parameter space. By contrast, local distance-ladder measurements constrain the low-redshift expansion rate more directly~\cite{HST:2000azd,Riess:2016jrr,Riess:2021jrx}.

Distance-based observables are intrinsically sensitive to integrated expansion histories. Since luminosity and angular-diameter distances involve integrals over $H(z)$, localized or time-varying fluctuations in the expansion rate can be partially smoothed through projection effects~\cite{Huterer:1998qv,Maor:2000jy,Astier:2000as,Huterer:2000mj,Huterer:2004ch,Linder:2005in,Wang:2005yaa,Amendola:2007rr,Lee:2025jrr}. As a result, multiple cosmological parameters may appear to be constrained even when the underlying information is effectively confined to a lower-dimensional subspace of parameter space~\cite{Bassett:2004wz,Efstathiou:1998xx,Lee:2025jrr}. This perspective allows us to identify how the increased precision in contemporary data, or curvature injection, can create a geometric wall that fundamentally limits the available pathways for tension reduction—a phenomenon we term \emph{geometric collision}. Related considerations have been emphasized in broader analyses of dataset tension, which argue that statistical criteria for identifying genuine signatures of new physics must account carefully for the structure of parameter inference and cross-probe comparisons~\cite{Cortes:2023dij,Cortes:2024lgw,Cortes:2025joz}. While these studies focus on the statistical interpretation of dataset tensions and the conditions under which a reduction of tension may legitimately support new physics, the present work addresses a complementary question. Rather than treating tension primarily as a scalar measure or model-selection diagnostic, we investigate its geometric structure in cosmological parameter space. In particular, we analyze how the discrepancy between early- and late-Universe probes can be decomposed into directional displacements of preferred parameters and directional curvatures supplied by individual experiments. This geometric perspective allows us to identify which parameter combinations control the rigidity of the joint constraints and how these directions are redistributed when the cosmological parameter space is enlarged.

In this work, we investigate the Hubble tension directly within the parameter space, decomposing it into two fundamental components: a displacement between the best-fit points and the directional curvature,  which quantifies the statistical penalty imposed by the data. Within the Gaussian approximation~\cite{Tegmark:1996bz,Albrecht:2006um}, the quadratic tension along a chosen direction factorizes into the squared parameter shift and the combined curvature supplied by the contributing probes~\cite{Verde:2013wza,Raveri:2015maa}. This formulation elucidates how model extensions can modify the inferred tension by altering the geometry of the constraint manifold.

We apply this decomposition to the $w$CDM model, in which the dark energy equation of state parameter is promoted from $w=-1$ to a free constant~\cite{Planck:2015fie,Planck:2018vyg,DiValentino:2021izs}. Using Planck, DESI DR2, and SH0ES~\cite{Planck:2018vyg,Riess:2011yx,Riess:2019cxk,Riess:2020fzl,DESI:2025zgx}, we examine how the dominant acoustic-scale constraint from the CMB~\cite{Planck:2018vyg,Hu:1995en} is redistributed when the parameter space is augmented, how curvature contributions combine across probes, and how the treatment of the sound horizon (rdFREE versus rdFIX) provides a controlled test of multi-probe geometry~\cite{Eisenstein:1997ik,BOSS:2014hhw,Brieden:2022heh}.

Our objective is not to advocate for a specific resolution to the tension, but to identify the geometric conditions under which a reduction in the nominal $H_0$ tension signifies genuine information gain as opposed to a mere redistribution of Fisher curvature within an augmented parameter space~\cite{Raveri:2018wln,Handley:2019wlz,Knox:2019rjx}. By distinguishing shift effects from curvature effects, we provide a transparent framework for interpreting model-dependent variations in the Hubble tension.

The remainder of this paper is organized as follows. Section~\ref{sec:parameter_space} formulates a parameter-space description of the Hubble tension, introducing the Gaussian approximation, the Fisher matrix formalism, and the separation between shift and curvature contributions. Section~\ref{sec:planck_geometry} reconstructs the Planck Fisher geometry and analyzes the statistical origin of the DESI DR2 parameter shifts, identifying the dominant acoustic-scale eigenmode and examining how curvature suppression and eigenmode rotation arise under model extension. Section~\ref{sec:mode_rotation} projects the Planck Fisher curvature into the $(\Omo, \ln H_0, w)$ space, distinguishing reparameterization effects from genuine curvature redistribution. In Section~\ref{sec:shift_curvature}, we develop the shift–curvature decomposition and introduce the notion of effective dimensionality. Section~\ref{sec:three_probe} extends the analysis to the full three-probe combination (Planck, DESI DR2, SH0ES), where we quantify curvature additivity, contrast the rdFREE and rdFIX treatments of the sound horizon, and isolate the intrinsic axial curvature supplied by SH0ES. Section~\ref{sec:discussion} discusses the geometric conditions under which model extensions can alleviate or preserve the $H_0$ tension. Concluding remarks are provided in Section~\ref{sec:conclusion}. Technical derivations and reproducibility details are provided in the appendices.

\section{Parameter-Space Description of the Hubble Tension}
\label{sec:parameter_space}
The statistical significance of the Hubble tension is determined not only by the relative shifts between parameter estimates inferred from early- and late-time probes, but also by the stiffness of the constraints along the relevant parameter combinations. To make this structure explicit, we formalize the tension directly within the parameter space.

\subsection{Gaussian approximation and Fisher matrix}
\label{subsec:21}
In the vicinity of the likelihood maximum, cosmological constraints are often well described by a local Gaussian approximation. For a parameter vector $\boldsymbol{\theta}$, the log-likelihood of a given dataset can be expanded as
\begin{equation}
-2\ln\mathcal{L} \simeq (\boldsymbol{\theta}-\boldsymbol{\theta}_\ast)^{\rm T} \,\mathbf{F}\, (\boldsymbol{\theta}-\boldsymbol{\theta}_\ast),
\end{equation}
where $\boldsymbol{\theta}_\ast$ denotes the best-fit point and $\mathbf{F}$ is the Fisher matrix, defined as
\begin{equation}
F_{ij} = -\left\langle \frac{\partial^2 \ln \mathcal{L}} {\partial \theta_i \partial \theta_j} \right\rangle .
\end{equation}
In this approximation, $\mathbf{F}$ defines the local curvature of the likelihood surface and serves as a Riemannian metric on the manifold of cosmological parameters. This information-geometric structure is typically hierarchical; it consists of stiff directions associated with large eigenvalues and flat, or sloppy, directions corresponding to weakly constrained parameter combinations.

\subsection{Shift and directional curvature}
\label{subsec:22}
Consider two statistically independent datasets with best-fit points $\boldsymbol{\theta}_1$ and $\boldsymbol{\theta}_2$. Their difference defines a parameter displacement vector
\begin{equation}
\Delta\boldsymbol{\theta} = \boldsymbol{\theta}_1-\boldsymbol{\theta}_2 .
\end{equation}
For a unit vector $\hat{\mathbf u}$ in parameter space, the projected shift along that direction is given by
\begin{equation}
\Delta_\parallel = \Delta\boldsymbol{\theta}\cdot\hat{\mathbf u}.
\end{equation}
The stiffness of the constraint along this direction is quantified by the Fisher matrix through
\begin{equation}
\kappa(\hat{\mathbf u}) = \hat{\mathbf u}^{\rm T} \,\mathbf{F}\, \hat{\mathbf u},
\end{equation}
which measures the directional curvature of the likelihood along $\hat{\mathbf u}$. We refer to $\Delta_\parallel$ as the \emph{projected shift} and to $\kappa(\hat{\mathbf u})$ as the \emph{directional curvature}.

\subsection{Quadratic tension}
\label{subsec:23}
Within the Gaussian approximation, the quadratic contribution to the tension along the direction $\hat{\mathbf u}$ is
\begin{equation}
T^2(\hat{\mathbf u}) = \kappa_{\rm joint}(\hat{\mathbf u}) \, \Delta_\parallel^2, \label{eq:tension_factorization}
\end{equation}
where $\kappa_{\rm joint}$ denotes the directional curvature of the combined datasets. Statistically, $T^2(\hat{\mathbf u})$ represents the Mahalanobis distance between the two distributions projected along the unit vector $\hat{\mathbf u}$. Equation~(\ref{eq:tension_factorization}) factorizes the tension into two distinct components: the magnitude of the displacement and the stiffness weighting of that displacement. This quadratic factorization provides the local geometric basis for several Bayesian tension metrics, such as the suspiciousness $S$ and the model-dimensionality $d_G$ \cite{Handley:2019wlz, Ong:2026tta}. In this framework, the tension is not merely a distance between means but a measure of how the Fisher information metric (curvature) penalizes any displacement away from the joint maximum-likelihood point.

\subsection{Additivity of curvature}
\label{subsec:24}
For statistically independent datasets, the Fisher information is additive,
\begin{equation}
\mathbf{F}_{\rm joint} = \sum_a \mathbf{F}_a .
\end{equation}
The joint directional curvature satisfies
\begin{equation}
\kappa_{\rm joint}(\hat{\mathbf u}) = \sum_a \hat{\mathbf u}^{\rm T} \,\mathbf{F}_a\, \hat{\mathbf u}.
\end{equation}
This relation allows the contribution of each probe to the total stiffness along a given direction to be identified directly.

In the following sections, we apply this parameter-space decomposition to the Planck, DESI DR2, and SH0ES constraints, examining how model extensions modify both the parameter displacements and the directional curvatures that enter the Hubble tension.

\section{Planck Fisher Geometry and the Statistical Origin of the DESI DR2 Parameter Shifts}
\label{sec:planck_geometry}
We investigate how extending the baseline cosmological model from $\Lambda$CDM to $\wCDM$ modifies the Fisher curvature of the Planck likelihood. A key objective is to determine whether the additional degree of freedom introduces a new high-curvature direction in parameter space or primarily suppresses and redistributes the curvature associated with the acoustic-scale constraint.

Throughout this section, we analyze the Fisher matrix reconstructed from the Planck 2018 \path{base_plikHM_TT_lowl_lowE} chains. For $\Lambda$CDM, we operate within the reduced subspace $(\Omo, \ln(h r_d))$, while for $\wCDM$, we consider the extended space $(\Omo, \ln(h r_d), w)$. This choice isolates the late-time directions most directly correlated with the CMB acoustic constraint.

\subsection{Acoustic-scale curvature in $\Lambda$CDM}
\label{subsec:31}
In $\Lambda$CDM, the dominant information from Planck arises from the acoustic angular scale,
\begin{equation}
\theta_* = \frac{r_s(z_*)}{D_A(z_*)}, \label{thetaast}
\end{equation}
where $r_s(z_*)$ is the comoving sound horizon at photon decoupling and $D_A(z_*)$ is the angular-diameter distance to that epoch. After marginalizing over early-universe parameters, the stringent constraint on $\theta_*$ induces a correlated variation between $\Omo$ and $H_0$. In the reduced late-time subspace, this manifests as a highly curved direction in the Fisher metric. Denoting the dominant eigenvector of the reconstructed Fisher matrix by
\begin{equation}
\mathbf v_1^{(\Lambda)} = (a_1,\, b_1),
\end{equation}
with components in the $(\Omo, \ln(h r_d))$ basis, small displacements along this direction approximately preserve the acoustic scale:
\begin{equation}
\delta \ln \theta_* \simeq a_1\,\delta \Omo + b_1\,\delta \ln(h r_d) \simeq 0.
\end{equation}

From the reconstructed Fisher matrix, we obtain
\begin{equation}
\mathbf v_1^{(\Lambda)} \simeq (0.0548,\;0.9985),
\end{equation}
indicating near-total alignment with the $\ln(h r_d)$ axis. The eigenvalue spectrum in this subspace is
\begin{equation}
(\lambda_1^{(\Lambda)},\lambda_2^{(\Lambda)}) \simeq (2.41\times10^{5},\,3.69\times10^{4}),
\end{equation}
yielding a hierarchy of
\begin{equation}
\lambda_1^{(\Lambda)} / \lambda_2^{(\Lambda)} \simeq 6.5.
\end{equation}
The leading eigenvalue corresponds to a sharply curved principal direction, while the subdominant eigenvalue reflects a comparatively flatter degeneracy manifold. The narrow Planck posterior, $\Omo = 0.3169 \pm 0.0065$ and $H_0 = 67.14 \pm 0.47$, is thus a projection of this acoustic-scale curvature rather than a direct measurement of the local expansion rate.

\subsection{Extension to $\wCDM$: curvature suppression and rotation}
\label{subsec:32}
Allowing the dark energy equation of state $w$ to deviate from $-1$ enlarges the parameter space and reconfigures the local Fisher geometry. Linearizing the acoustic condition in the extended space $(\Omo, \ln(h r_d), w)$ yields
\begin{equation}
\delta \ln \theta_* \simeq a_1\,\delta\Omo + b_1\,\delta\ln(h r_d) + c_1\,\delta w \simeq 0,
\end{equation}
where $(a_1,b_1,c_1)$ are the components of the dominant Fisher eigenvector.

Diagonalizing the reconstructed Planck Fisher matrix in $\wCDM$ yields
\begin{equation}
\lambda_1^{(w)} \simeq 6.44\times10^{3}, \qquad \mathbf v_1^{(w)} \simeq (0.9997,\;0.00326,\;-0.0244).
\end{equation}
Two distinct geometric effects relative to $\Lambda$CDM are evident. First, the leading eigenvalue is suppressed by nearly two orders of magnitude:
\begin{equation}
\lambda_1^{(w)} / \lambda_1^{(\Lambda)} \simeq 2.7\times10^{-2}.
\end{equation}
This drastic reduction in principal stiffness directly accounts for the broadening of the Planck-only posterior in $\wCDM$. Second, to quantify the change in orientation (\textit{eigenmode rotation}), we compare the dominant eigenvector in $\wCDM$ with its $\Lambda$CDM counterpart. Projecting $\mathbf v_1^{(w)}$ onto the $(\Omo, \ln(h r_d))$ plane, we define
\begin{equation}
\widetilde{\mathbf v}_1^{(w)} = \frac{(a_1^{(w)},\,b_1^{(w)})}{\sqrt{(a_1^{(w)})^2+(b_1^{(w)})^2}},
\end{equation}
where the rotation angle $\phi_{\rm ac}$ relative to $\mathbf v_1^{(\Lambda)}$ is given by
\begin{equation}
\cos \phi_{\rm ac} = \widetilde{\mathbf v}_1^{(w)} \cdot \mathbf v_1^{(\Lambda)} .
\end{equation}
Evaluating this gives
\begin{equation}
\phi_{\rm ac} \simeq 86.7^\circ .
\end{equation}
This near-orthogonal rotation demonstrates that the extension to $\wCDM$ does not merely add a new parameter; it effectively dilutes the acoustic curvature that constrained $(\Omo, H_0)$ by rotating the principal stiffness direction away from the $\ln(h r_d)$ axis.

Physically, this curvature suppression accounts for the expansion of the posterior volume, which in a Bayesian context manifests as an Ockham penalty. The factor of $\sim 10^{-2}$ reduction in the leading eigenvalue implies a substantial dilution of information along the primary acoustic-scale direction. This provides a physical explanation for why recent Bayesian analyses find that the inclusion of $w$ as a free parameter fails to improve the model evidence despite the lower $\chi^2$ values \cite{Ong:2026tta, Ormondroyd:2025phk}; the gain in the goodness-of-fit is statistically outweighed by the loss of geometric stiffness.

\subsection{DESI DR2 as curvature injection}
\label{subsec:33}
The posterior values reported in DESI DR2~\cite{DESI:2025zgx} corroborate the Fisher-geometric structure derived above. For Planck alone in $\wCDM$:
\begin{equation}
\Omo = 0.203^{+0.017}_{-0.060}, \qquad H_0 = 85^{+10}_{-6}, \qquad w = -1.55^{+0.17}_{-0.37}, \label{PwCDM}
\end{equation}
whereas the combined Planck+DESI constraint yields:
\begin{equation}
\Omo = 0.2927 \pm 0.0073, \qquad H_0 = 69.51 \pm 0.92, \qquad w = -1.055 \pm 0.036. \label{PDESIwCDM}
\end{equation}
In $\wCDM$, the suppression and rotation of the acoustic-aligned eigenmode weaken the effective curvature in the $(\Omo, H_0)$ plane, leading to the broad marginal posterior in Eq.~(\ref{PwCDM}). The large excursion toward low $\Omo$ and high $H_0$ is driven not by a new stiff direction along $w$, but by the dilution of the pre-existing acoustic curvature.

DESI DR2, which constrains late-time distance ratios $D_M(z)/r_d$ and $D_H(z)/r_d$, is primarily sensitive to the combination $h r_d$. In Fisher terms, DESI contributes curvature in a direction with a substantial projection onto the suppressed Planck acoustic manifold. Combining the datasets, $\mathbf F_{\rm joint} = \mathbf F_{\rm Planck} + \mathbf F_{\rm DESI}$, the previously shallow eigenmode in $\wCDM$ acquires additional stiffness. The posterior contraction from Eq.~(\ref{PwCDM}) to Eq.~(\ref{PDESIwCDM}) is a direct manifestation of \emph{curvature injection}.

In contrast, the dominant Planck eigenmode in $\Lambda$CDM is already well aligned with $\ln(h r_d)$. The Planck+DESI result,
\begin{equation}
\Omo = 0.3027 \pm 0.0036, \qquad H_0 = 68.17 \pm 0.28, \label{PDESILCDM}
\end{equation}
shows only a mild shift relative to Planck alone, consistent with an enhancement of the leading eigenvalue without substantial eigenvector rotation. Thus, DESI stabilizes a previously diluted acoustic mode in $\wCDM$, whereas it simply reinforces an already stiff one in $\Lambda$CDM. This stabilization via curvature injection is precisely what restricts the phantom escape in joint analyses. The high precision of DESI DR2 essentially re-establishes the geometric wall that $\Lambda$CDM originally imposed, a phenomenon captured by the localized tension signatures (low $d_G$) in global Bayesian fits \cite{Ong:2026tta}.

\subsection{Model-dependent curvature structure and parameter inference}
\label{subsec34}
The Fisher analysis demonstrates that the Planck-only behavior in $\wCDM$ is primarily a geometric effect. The broad posterior for $H_0$ and the associated shifts in $(\Omo, H_0)$ do not necessarily imply a statistical preference for $w \neq -1$; rather, they result from the reorganization of likelihood curvature within the augmented parameter space.

From this perspective, the impact of DESI DR2 is interpreted through curvature additivity. In $\wCDM$, where the Planck curvature is reorganized, the DESI contribution modifies the joint eigenstructure more visibly than in $\Lambda$CDM, where the dominant curvature direction remains fixed. These results are reconstructed directly from the Planck 2018 chains, and we have verified that the leading eigenvalue hierarchy is driven by the data rather than prior truncation. Our reconstruction confirms that the reported shifts in $H_0$ under $w$CDM are not signatures of new physics, but rather predictable consequences of the reorganization of the Fisher information metric within the augmented parameter space.

\section{Projection of the Planck Fisher Curvature into $(\Omo, \ln H_0, w)$ Space}
\label{sec:mode_rotation}
In Sec.~\ref{sec:planck_geometry}, we analyzed the Planck likelihood in the acoustic-aligned basis $(\Omo,\ln(hr_d))$, where the dominant Fisher eigenmode is directly associated with preservation of the acoustic angular scale $\theta_\ast$. In that representation, the leading eigenvalue quantified the intrinsic acoustic rigidity of the Planck constraint and clarified the geometric origin of its strong anisotropy.

In the present section, we reformulate this structure in terms of the expansion-rate variable $H_0$. While the numerical analysis is performed in the $(\Omo,\ln H_0)$ parameter space, the physical interpretation will be presented in terms of both $H_0$ and $\ln H_0$. The logarithmic parameterization is adopted for numerical stability and for convenience in the Fisher analysis: derivatives with respect to $\ln H_0$ correspond to fractional variations of the Hubble scale,
\begin{equation}
\frac{\partial}{\partial \ln H_0} = H_0 \frac{\partial}{\partial H_0},
\end{equation}
so that the Fisher curvature directly measures the sensitivity of the likelihood to relative changes in the expansion rate. This parameterization also avoids scale-dependent conditioning of the Fisher matrix and leads to more stable numerical reconstruction of the covariance from the MCMC chains.

We now examine how the Planck Fisher curvature behaves when expressed in this physically transparent expansion-rate direction and how it is modified when the cosmological model is extended. Two distinct operations must be separated. First, one may perform a reparameterization within a fixed cosmological model. Second, one may extend the model itself, for example, from $\Lambda$CDM to $w$CDM. The former does not alter the likelihood hypersurface; it merely redistributes the same Fisher curvature under a coordinate transformation. The latter changes the hypersurface and can modify both the curvature scale and the dimensionality of the degeneracy surface.

This distinction is essential. If an apparent change in the Planck $H_0$ constraint were purely a coordinate effect, it would carry no physical implication. By contrast, a genuine model extension may weaken the effective curvature scale or enlarge the acoustic-preserving manifold in parameter space.

Within $\Lambda$CDM, transforming from $\ln(hr_d)$ to $\ln H_0$ is a change of coordinates. The likelihood itself remains invariant,
\begin{equation}
\mathcal{L}_{\Lambda}(\Omo,\ln(hr_d)) = \mathcal{L}_{\Lambda}(\Omo,\ln H_0), \label{LLCDM}
\end{equation}
while the Fisher matrix transforms covariantly,
\begin{equation}
\mathbf{F}'_{\Lambda} = \mathbf{J}^{\rm T} \mathbf{F}_{\Lambda} \mathbf{J} ,\label{FLCDM}
\end{equation}
where $\mathbf{J}=\partial\theta/\partial\theta'$ is the Jacobian evaluated at the maximum-likelihood point. Because the Fisher matrix corresponds to the second derivative of $\ln\mathcal{L}$, linear reparameterizations preserve its eigenvalue spectrum. In $\Lambda$CDM, the mapping between $\ln(hr_d)$ and $\ln H_0$ is effectively linear, since the sound horizon is fixed by early-time physics. Consequently, the curvature amplitudes remain unchanged, although their projections onto coordinate axes are redistributed. Any apparent sharpening of the Planck constraint in $H_0$-space within $\Lambda$CDM therefore reflects geometric projection of the same acoustic rigidity identified in Sec.~\ref{sec:planck_geometry}.

The situation differs when the model is extended. Allowing $w$ to vary introduces an additional physical degree of freedom, so that
\begin{equation}
\mathcal{L}_{\Lambda}(\Omo, \ln H_0) \neq \mathcal{L}_{w}(\Omo, \ln H_0, w), \label{LwCDM}
\end{equation}
and the Fisher tensor becomes
\begin{equation}
\mathbf{F}_{w} = - \left\langle \frac{\partial^2 \ln \mathcal{L}_{w}}{\partial \theta_i \partial \theta_j} \right\rangle . \label{FwCDM}
\end{equation}
In this case, both the curvature amplitudes and the effective dimensionality of the acoustic-degeneracy surface may change. The correlated manifold defined by $\theta_\ast=\text{const}$, which is effectively two-dimensional in $\Lambda$CDM, becomes embedded in a higher-dimensional surface when $w$ is promoted to a free parameter.

\subsection{Reparameterization within $\Lambda$CDM}
\label{subsec:41}
Within $\Lambda$CDM, transforming from $\ln(hr_d)$ to $\ln H_0$ constitutes a linear reparameterization of the late-time sector. Under such linear transformations, the eigenvalue spectrum of the Fisher matrix is invariant, although the orientation of the eigenvectors with respect to coordinate axes is redistributed.

Projecting the reconstructed Planck Fisher tensor onto the two-dimensional $(\Omo,\ln H_0)$ subspace and diagonalizing yields
\begin{equation}
(\lambda_1^{(\Lambda)},\lambda_2^{(\Lambda)}) \simeq (7.50\times10^{5},\,1.15\times10^{5}) .
\end{equation}
The leading curvature scale $\lambda_1^{(\Lambda)} \simeq 7.50\times10^{5}$ signals a highly rigid local direction in the $(\Omo,\ln H_0)$ plane. The corresponding eigenvector $\mathbf{v}_1$ is misaligned with the pure $\ln H_0$ axis by an angle $\theta_{\Lambda} = \cos^{-1}(|\mathbf{v}_1 \cdot \hat{\mathbf{e}}_{\ln H_0}|) \simeq 46.4^\circ$. This tilt is the projection of the acoustic-aligned eigenmode. The eigenvalue hierarchy $\lambda_1^{(\Lambda)} / \lambda_2^{(\Lambda)} \simeq 6.52$ confirms that the anisotropy of the Fisher metric is basis invariant within $\Lambda$CDM.

\subsection{Fisher curvature redistribution in $w$CDM}
\label{subsec:wCDM}
The extension to $(\Omo,\ln H_0, w)$ introduces an additional physical degree of freedom. Diagonalizing the reconstructed three-dimensional Fisher tensor yields
\begin{equation}
(\lambda_1^{(w)},\lambda_2^{(w)},\lambda_3^{(w)}) \simeq (2.00\times10^{4},\,1.72\times10^{3},\,9.29) .
\end{equation}
Relative to the $\Lambda$CDM projection, the leading eigenvalue is strongly suppressed: $\lambda_1^{(\Lambda)} / \lambda_1^{(w)} \simeq 37.5$. This corresponds to an increase in the characteristic uncertainty width along the principal direction by a factor of $\sqrt{37.5}\simeq 6.1$.

\begin{figure}[t]
\centering
\includegraphics[width=0.78\linewidth]{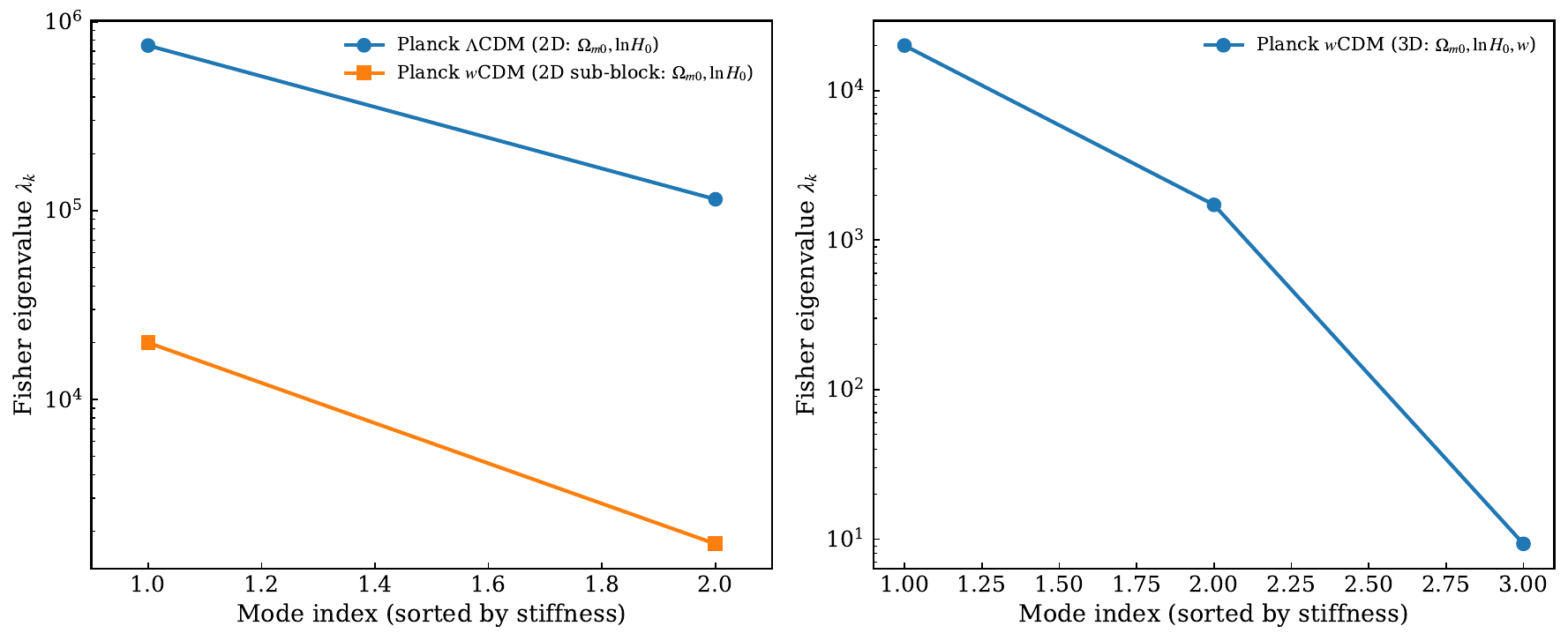}
\caption{Redistribution of the Planck Fisher eigenvalue spectrum under the $\Lambda$CDM $\to$ $w$CDM extension. \textbf{Left}: Comparison of leading eigenvalues. $\Lambda$CDM (filled circles) and $w$CDM (filled squares) illustrate the principal stiffness suppression by a factor $\sim 37.5$. \textbf{Right}: Full three-dimensional spectrum in $w$CDM (filled circles), where the third eigenvalue indicates the emergence of a new degeneracy direction. The Fisher curvature is redistributed, weakening the acoustic rigidity without introducing a new stiff direction aligned with $w$. }
\label{fig:1}
\end{figure}

Figure~\ref{fig:1} visualizes this redistribution. Left panel compares the leading eigenvalues in the $(\Omo, \ln H_0)$ projection, where $\Lambda$CDM (filled circles connected by a solid line) and $w$CDM (filled squares connected by a solid line) results are shown. The vertical offset quantifies the weakening of the effective acoustic rigidity. Right panel displays the full three-dimensional spectrum in $w$CDM space (filled circles). In addition to the suppressed leading mode, a highly subdominant third eigenvalue $\lambda_3^{(w)}$ appears, reflecting the extension of the condition $\theta_\ast=\text{const}$ into the additional parameter dimension. Thus, the model extension dilutes the pre-existing acoustic constraint across an enlarged parameter manifold rather than generating new independent information. This suppression factor of $\sim 37.5$ provides a precise information-geometric quantification of the Bayesian Ockham penalty reported in recent model-selection studies \cite{Ong:2026tta, Ormondroyd:2025phk}. The dramatic flattening of the likelihood surface along the principal acoustic-preserving axis increases the effective parameter volume, explaining why the Bayesian evidence remains low despite the model's increased flexibility to accommodate higher $H_0$ values.

\subsection{Curvature magnitude versus orientation}
\label{subsec:curvature_projection}
Despite the suppression of the leading eigenvalue magnitude, the orientation of the principal constraint direction remains nearly stable. Projecting the leading eigenvectors onto the $(\Omo, \ln H_0)$ subspace, we find $\cos\phi_{H_0} \simeq 0.988$, implying a rotation of only $\phi_{H_0} \simeq 9.06^\circ$. This indicates that the principal constraint remains closely aligned with the acoustic-preserving combination. Geometrically, the condition $\theta_\ast=\text{const}$ becomes embedded in a higher-dimensional degeneracy surface, but its projection onto the late-time plane remains nearly parallel to the original direction. This geometric stability implies that the phantom escape route ($w < -1$) identified in our $w$-scan is not a result of a new physical direction being opened, but rather a consequence of the pre-existing acoustic manifold becoming softer. Consequently, any shift along this path is statistically expensive in a joint analysis, as it must compete with the curvature injection from late-time probes that reinforce the $\Lambda$CDM-like rigidity.

\subsection{Validation of the Local Gaussian Approximation}
\label{subsec:gaussian_validation}
\begin{figure}[t]
\centering
\includegraphics[width=0.78\linewidth]{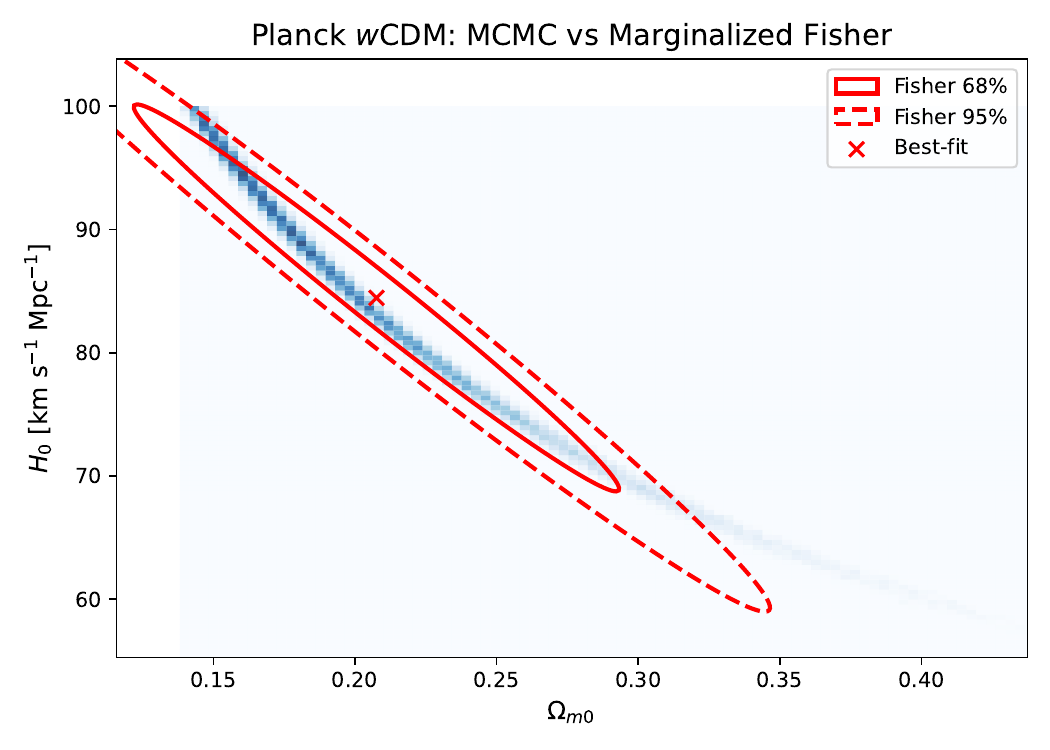}
\caption{\textbf{Validation of the local Gaussian approximation in Planck $w$CDM.} The shaded density map shows the full MCMC posterior in the $(\Omo, H_0)$ plane. Solid and dashed ellipses denote the 68\% and 95\% confidence regions derived from the marginalized Fisher matrix, while the cross marks the maximum-likelihood point. The Fisher ellipse faithfully reproduces the local orientation and curvature scale near the maximum, justifying its use in the present analysis.}
\label{fig:planck_wcdm_gaussian_validation}
\end{figure}

We verify the accuracy of the local stiffness structure through a comparison with the full MCMC chains. Figure~\ref{fig:planck_wcdm_gaussian_validation} overlays the marginalized Fisher ellipses on the posterior density map.

The shaded density map represents the full Planck $w$CDM posterior, while the solid and dashed ellipses indicate the 68\% and 95\% confidence regions derived from the Fisher matrix. Although the global distribution exhibits a mildly curved degeneracy, the Fisher ellipse closely reproduces the local orientation and curvature scale near the maximum-likelihood point. This demonstrates that the Fisher tensor provides a sufficient representation of the local directional curvature relevant for the tension analysis. The faithfulness of the local Gaussian representation justifies our use of the Fisher-geometric identity $T^2 = \kappa \Delta^2$. It ensures that the decomposed contributions of shift and stiffness accurately reflect the underlying likelihood structure, even when the global posterior exhibits mild non-Gaussian degeneracies.

\section{Shift–Curvature Decomposition of the Planck Geometry}
\label{sec:shift_curvature}

Section~\ref{sec:mode_rotation} showed that extending $\Lambda$CDM to $w$CDM leaves the principal Planck constraint direction nearly unchanged while strongly suppressing the leading Fisher eigenvalue (Eq.~\eqref{SuppressionPlanck}). We now examine how this curvature suppression enters the statistical interpretation of parameter discrepancies, focusing in particular on the Hubble tension.

Within the local Gaussian approximation, the Fisher matrix defines the metric structure governing statistical penalties in parameter space.  In this section, we express the parameter discrepancy between probes as a displacement measured in the Planck Fisher geometry and analyze how the associated quadratic penalty depends on the curvature structure derived in the previous section.

\subsection{Quadratic tension and statistical interpretation}
\label{subsec:51}
To connect the geometric results of Secs.~\ref{sec:planck_geometry} and~\ref{sec:mode_rotation} with statistical tension measures, we introduce a curvature-weighted displacement defined in the Planck Fisher metric. Let $\Delta\boldsymbol{\theta}$ denote the parameter displacement between the best-fit point preferred by Planck and that preferred by an external probe. The Planck-metric quadratic form is
\begin{equation}
T_P^2 \equiv \Delta\boldsymbol{\theta}^{\mathsf T} \mathbf F_P \Delta\boldsymbol{\theta}, \label{eq:T2_def}
\end{equation}
where $\mathbf F_P$ is the Planck Fisher matrix.  This quantity does not represent a symmetric two-probe tension statistic; instead, it measures how strongly a given parameter displacement is penalized by the local curvature of the Planck likelihood. In the Gaussian approximation, covariance-weighted quadratic forms of this type are widely used in cosmological tension analyses~\cite{Verde:2013wza,Lin:2017ikq,Raveri:2018wln,Handley:2019wlz}.  The symmetric multi-probe Gaussian derivation employed for the numerical calculations in this work is presented in Appendix~\ref{app:complete_square_multprobe}.

Using the spectral decomposition of the Planck Fisher tensor,
\begin{equation}
\mathbf F_P = \sum_\alpha \lambda_\alpha \mathbf e_\alpha \mathbf e_\alpha^{\mathsf T}, \label{eq:F_spectral}
\end{equation}
the quadratic form becomes
\begin{equation}
T_P^2 = \sum_\alpha \lambda_\alpha (\Delta_\alpha)^2, \qquad \Delta_\alpha = \mathbf e_\alpha^{\mathsf T}  \Delta\boldsymbol{\theta}. \label{eq:T2_modes}
\end{equation}
Equation~(\ref{eq:T2_modes}) shows that the statistical penalty depends on both the displacement along each principal eigendirection and the corresponding curvature amplitude. A reduction in $T_P^2$ may therefore arise either from a smaller parameter shift or from suppression of the relevant Fisher eigenvalue.

For a unit vector $\hat{\mathbf u}$ in parameter space, the directional curvature of the Planck metric is
\begin{equation}
\kappa_P(\hat{\mathbf u}) = \hat{\mathbf u}^{\mathsf T} \mathbf F_P \hat{\mathbf u} = \sum_\alpha \lambda_\alpha
(\mathbf e_\alpha \cdot \hat{\mathbf u})^2 . \label{eq:kappa_general}
\end{equation}
Projecting the displacement vector onto $\hat{\mathbf u}$,
\begin{equation}
\Delta\boldsymbol{\theta}_{\parallel} = (\hat{\mathbf u}^{\mathsf T} \Delta\boldsymbol{\theta}) \,\hat{\mathbf u},
\end{equation}
gives the projected quadratic tension
\begin{equation}
T_{P,\hat{\mathbf u}}^2 = \kappa_P(\hat{\mathbf u}) (\Delta u)^2 \,, \qquad \Delta u = \hat{\mathbf u}^{\mathsf T}
\Delta\boldsymbol{\theta}. \label{eq:TP_dir}
\end{equation}
Unless $\hat{\mathbf u}$ coincides with a Fisher eigenvector, this expression represents a one-dimensional projection of the full quadratic form.

For the expansion-rate direction $\hat{\mathbf u}_{\ln H_0}$,
\begin{equation}
\kappa_P(\ln H_0) = \hat{\mathbf u}_{\ln H_0}^{\mathsf T} \mathbf F_P \hat{\mathbf u}_{\ln H_0}, \label{eq:kappa_lnH0}
\end{equation}
and the projected quadratic tension becomes
\begin{equation}
T_{P,\ln H_0}^2 = \kappa_P(\ln H_0) (\Delta\ln H_0)^2 . \label{eq:T2_dir}
\end{equation}
Thus, for a fixed displacement amplitude $\Delta\ln H_0$, the effective tension decreases whenever the directional curvature $\kappa_P(\ln H_0)$ is suppressed.

In the Gaussian approximation, the Fisher matrix is the inverse covariance, $\mathbf F_P=\mathbf C_P^{-1}$, so the quadratic form behaves as a $\chi^2_k$ statistic if the posterior is locally Gaussian. For the one-dimensional projection along $\ln H_0$,
\begin{equation}
T_{P,\ln H_0} = \sqrt{\kappa_P(\ln H_0)} |\Delta\ln H_0| ,. 
\end{equation}
A $3\sigma$ discrepancy corresponds to
\begin{equation}
T_{P,\ln H_0}^2 = 9.
\end{equation}

Because the $w$CDM extension leaves the principal eigendirection nearly unchanged while suppressing the leading eigenvalue, the stiffness entering Eq.~(\ref{eq:T2_dir}) inherits this suppression. The reduction of the Planck-only Hubble tension in $w$CDM therefore follows primarily from the weakening of the relevant curvature scale rather than from a reorientation of the dominant constraint direction. This factorization reveals that the apparent alleviation of the Hubble tension in $w$CDM is primarily a consequence of information dilution (lower $\kappa_P$) rather than a physical convergence of the preferred $H_0$ values. From this perspective, the geometric framework allows for a direct diagnostic of the tension's origin without requiring the exhaustive posterior sampling typically employed in Bayesian model-selection studies \cite{Ong:2026tta}.

\subsection{Curvature suppression and effective dimensionality}
\label{subsec:52}
The eigenmode decomposition introduced above clarifies not only how curvature amplitudes enter the quadratic tension, but also how many independent curvature modes contribute significantly to the stiffness along a chosen parameter direction.

For the expansion-rate direction, we define
\begin{equation}
A_\alpha(\ln H_0) = \lambda_\alpha \left( \mathbf e_\alpha \cdot \hat{\mathbf u}_{\ln H_0} \right)^2 , \label{eq:Aalpha_def}
\end{equation}
which quantifies the contribution of eigenmode $\alpha$ to the directional curvature. Summing over all modes reproduces
\begin{equation}
\kappa_P(\ln H_0) = \sum_\alpha A_\alpha(\ln H_0). \label{eq:kappa_sum}
\end{equation}

To characterize how the directional stiffness builds up across the eigenmode spectrum, we introduce the cumulative curvature fraction
\begin{equation}
C_k(\ln H_0) = \frac{ \sum_{\alpha=1}^{k} A_\alpha(\ln H_0)}{\kappa_P(\ln H_0)}, \label{eq:Ck_def}
\end{equation}
where eigenmodes are ordered by descending $\lambda_\alpha$.  Therefore,  the function $C_k(\ln H_0)$ measures how rapidly the curvature along $\ln H_0$ is saturated as additional Fisher eigenmodes are included. If $C_1(\ln H_0)\simeq 1$, the stiffness along $\ln H_0$ is dominated by a single principal mode.  If instead $C_k$ grows gradually with $k$, several eigendirections contribute comparably to the directional curvature. For both $\Lambda$CDM and $w$CDM we find that the curvature along
$\ln H_0$ is concentrated in a very small number of leading Fisher modes. In the rdFIX configuration, the cumulative curvature is already close to single–mode dominance, with $C_1(\ln H_0)\simeq1$, while in the rdFREE configuration the stiffness is shared by the first two eigenmodes,so that the cumulative fraction saturates only at $C_2(\ln H_0)=1$.The strong suppression of the leading eigenvalue identified in Sec.~\ref{sec:mode_rotation} therefore reduces the magnitude of the directional curvature. While the dominant stiffness remains concentrated in a small set of leading eigenmodes, the relative contributions of the first two modes can vary depending on the treatment of the sound horizon. The high degree of curvature localization ($C_1 \simeq 1$ in rdFIX) provides a geometric explanation for the low effective dimensionality ($d_G \approx 1-3$) reported in recent Bayesian tension heatmaps \cite{Ong:2026tta}. While Bayesian methods arrive at this low dimensionality through high-dimensional integration, our $C_k$ analysis identifies it directly from the hierarchical eigenstructure of the Fisher information metric, confirming that the tension is dominated by a single stiff mode associated with the acoustic scale.

From the statistical perspective of Eq.~(\ref{eq:T2_def}), this implies that the reduction of the Planck-only Hubble tension in $w$CDM does not arise from an increase in the number of statistically independent curvature directions. Instead, it follows directly from eigenvalue suppression: when the dominant curvature scale decreases, the quadratic penalty assigned to a fixed parameter displacement is correspondingly reduced. Geometrically, the Planck constraint hypersurface broadens while preserving nearly the same principal orientation.

Finally, we emphasize that the Planck-metric quantity $T_P^2$ is not itself a symmetric multi-probe tension statistic. In a Gaussian multi-probe combination, the Fisher matrices are additive,
\begin{equation}
F_{\rm joint} = F_P + F_X ,
\end{equation}
so that
\begin{equation}
T_{\rm joint}^2 = \Delta\theta^{\mathsf T} (F_P + F_X) \Delta\theta = T_P^2 + T_X^2 .
\end{equation}
Because each quadratic contribution is positive definite, tension cannot be reduced by cancellation of curvature penalties. Apparent reductions arise instead from either smaller parameter displacements or suppression of the relevant curvature scale within a given probe. The present analysis isolates the latter mechanism within the Planck Fisher geometry and provides the geometric basis for the multi-probe curvature attribution discussed in Sec.~\ref{sec:three_probe}.  This additivity implies that if a new probe (such as DESI DR2) injects curvature along an existing stiff direction, the joint tension $T_{\rm joint}^2$ must inevitably increase unless the parameter displacement $\Delta\theta$ is simultaneously reduced. Our framework thus characterizes the Hubble tension as a geometric collision between the rigid constraints of early-universe physics and the increasingly stiff observations from late-time probes. For clarity, the geometric diagnostics used throughout the present analysis are summarized in Table~\ref{tab:diagnostics_summary}.

\begin{table*}[t]
\centering
\caption{Summary of geometric diagnostics used in this work. The table lists the mathematical definitions, their geometric meaning in Fisher space, and their role in diagnosing the stiffness structure of cosmological parameter constraints. }
\label{tab:diagnostics_summary}
\begin{tabular}{l p{3.2cm} p{3.2cm} p{3.5cm} p{3.2cm}}
\hline\hline
Quantity & Definition & Geometric Meaning & Physical Interpretation & Diagnostic Role \\
\hline
$w_\alpha(\phi)$ & $|\hat{\mathbf e}_\phi \!\cdot\! \mathbf v_\alpha|^2$ & Alignment of direction $\phi$ with eigenmode $\alpha$ & Geometric overlap between parameter direction and eigenstructure &
Identifies which modes support the direction \\
$\lambda_\alpha w_\alpha(\phi)$ & Eigenvalue-weighted projection & Curvature contribution from eigenmode $\alpha$
& Fraction of stiffness along $\phi$ carried by that mode & Distinguishes stiff and sloppy support \\
$\kappa(\phi)$ & $\hat{\mathbf e}_\phi^{\mathsf T}\mathbf F\hat{\mathbf e}_\phi$ &
Directional curvature along $\phi$ & Total effective stiffness along the parameter direction &
Sets the quadratic tension scale
\\
$C_k(\phi)$ & $\displaystyle \sum_{\alpha=1}^{k} \lambda_\alpha w_\alpha(\phi) $ &
Cumulative curvature from leading $k$ modes & Build-up of stiffness from dominant eigenmodes &
Tracks curvature accumulation \\
$\displaystyle \frac{C_k(\phi)}{\kappa(\phi)}$ & Normalized cumulative curvature & Fraction of total curvature in first $k$ modes
& Localization of stiffness in eigenmode space & Diagnoses effective dimensionality \\
\hline\hline
\end{tabular}
\end{table*}
The diagnostics summarized in Table~\ref{tab:diagnostics_summary} provide a unified language for describing how curvature is distributed across parameter directions and Fisher eigenmodes. In the previous sections, we have applied these tools to the Planck Fisher geometry in isolation, showing that the $w$CDM extension suppresses the dominant curvature scale while leaving the principal constraint direction largely unchanged.

We now extend this geometric analysis to the joint multi-probe problem. In particular, we examine how the curvature structures of Planck, DESI DR2, and SH0ES combine within the common parameter space $(\Omo,\ln H_0,w)$, and how their directional projections determine the statistical significance of the Hubble tension. This multi-probe curvature attribution is developed in Sec.~\ref{sec:three_probe}.

\section{Three-Probe Curvature Attribution in $w$CDM}
\label{sec:three_probe}
Sections~\ref{sec:mode_rotation} and~\ref{sec:shift_curvature} established two key geometric properties of the Planck Fisher information in $w$CDM. First, extending $\Lambda$CDM to $w$CDM primarily suppresses the leading curvature amplitude while leaving the dominant eigendirection in the $(\Omo,\ln H_0)$ subspace nearly unchanged. Second, the directional curvature spectrum along $\ln H_0$ is strongly hierarchical, so that the effective dimensionality of the Planck constraint in expansion-rate space remains close to unity.

These results characterize the internal curvature structure of the Planck likelihood. The Hubble tension, however, is not determined by Planck alone. It arises from the combination of independent Fisher tensors contributed by different cosmological probes. Therefore, the relevant geometric quantity is the joint directional curvature along the physical expansion-rate axis,
\begin{equation}
T^2_{\ln H_0} = \kappa_{\rm joint}(\ln H_0) (\Delta \ln H_0)^2 ,
\end{equation}
as introduced in Sec.~\ref{sec:shift_curvature}. The displacement $\Delta\ln H_0$ represents the shift between probes, while $\kappa_{\rm joint}$ quantifies the stiffness of the combined Fisher geometry along that direction. Thus, understanding the Hubble tension requires identifying how directional curvature along the expansion rate axis is distributed across independent probes. 

In this section, we decompose the joint Fisher curvature into contributions from Planck, DESI DR2, and SH0ES, and analyze how this curvature attribution depends on the treatment of the sound horizon.

\subsection{Curvature additivity and directional projection}
\label{subsec:61}

We adopt the unified parameter basis $\boldsymbol{\theta}=(\Omo,\ln H_0, w)$, identical to that used in Sec.~\ref{sec:mode_rotation}. Under the local Gaussian approximation and neglecting cross-covariances between statistically independent experiments, Fisher information is additive,
\begin{equation}
\mathbf F_{\rm joint} = \mathbf F_P + \mathbf F_D + \mathbf F_S ,
\end{equation}
where $P$, $D$, and $S$ denote Planck (CMB acoustic scale), DESI DR2 (BAO geometry), and SH0ES (local distance ladder), respectively. Each experiment contributes an independent local curvature tensor in parameter space, and the joint Fisher geometry is obtained by the superposition of these curvature sources.

To quantify the stiffness along the physical expansion-rate axis, we project each Fisher tensor onto the unit vector $\hat{\mathbf u}_{\ln H_0}$,
\begin{equation}
\kappa_X(\ln H_0) = \hat{\mathbf u}_{\ln H_0}^{\mathsf T} \mathbf F_X \hat{\mathbf u}_{\ln H_0},
\end{equation}
where $X \in \{P,D,S\}$. The joint directional curvature is given by
\begin{equation}
\kappa_{\rm joint}(\ln H_0) = \kappa_P + \kappa_D + \kappa_S .
\end{equation}
The quadratic tension metric introduced in Sec.~\ref{sec:shift_curvature},
\begin{equation}
T^2_{H_0} = \kappa_{\rm joint}(H_0) (\Delta H_0)^2 ,
\label{T2H0}
\end{equation}
makes explicit that the statistical significance of the Hubble tension is controlled by two independent ingredients: the displacement $\Delta H_0$ and the curvature scale that weights this displacement. Unlike a direct comparison of best-fit values, this formulation separates parameter shift from directional stiffness within a unified information-geometric framework. From a Bayesian perspective, this combined stiffness $\kappa_{\rm joint}$ determines the tightness of the joint posterior; a higher curvature implies a higher statistical penalty for any displacement $\Delta H_0$. This explains why the inclusion of high-precision DESI DR2 data reinforces the $\Lambda$CDM model in global fits \cite{Ong:2026tta}---the injected curvature is simply too large to be offset by the model's added flexibility.

To characterize curvature attribution, we define fractional contributions
\begin{equation}
f_X = \frac{\kappa_X}{\kappa_{\rm joint}},
\end{equation}
which quantify how much of the joint rigidity along the expansion-rate axis is supplied by each probe.

Each experiment is originally defined in its own native parameterization. To ensure consistency of the directional projection, all Fisher tensors are first transformed into the common $(\Omo,\ln H_0, w)$ basis via the appropriate Jacobians (cf.~Sec.~\ref{sec:mode_rotation}). Only after this basis unification is the displacement vector $\Delta\boldsymbol{\theta}$ constructed and projected onto $\hat{\mathbf u}_{\ln H_0}$. All numerical results reported below are obtained from this Jacobian--consistent implementation (Appendix~\ref{app:complete_square_multprobe}), ensuring full reproducibility of the curvature fractions.

The three probes contribute to the joint curvature through structurally different information channels. Planck constrains the expansion rate indirectly through preservation of the acoustic angular scale $\theta_\ast$, which generates stiffness along a correlated acoustic-scale direction in parameter space. SH0ES determines $H_0$ directly in absolute physical units, producing intrinsic axial curvature aligned with the expansion-rate axis. DESI constrains combinations involving $h$ and $r_d$ through BAO distance ratios, so that its projected curvature along $\ln H_0$ depends sensitively on how the sound horizon is treated~\cite{Lee:2025rmg,Lee:2025ysg}.

Therefore,  the Hubble tension reflects a redistribution of curvature within the combined Fisher geometry. The same displacement $\Delta H_0$ can acquire different statistical weights depending on how directional stiffness is partitioned among these structurally distinct information sources.

\subsection{rdFREE: Basis Dependence of the Effective Two-Probe Geometry}
\label{subsec:62}

We now analyze the rdFREE configuration, in which the sound horizon $r_d$ is marginalized in the BAO likelihood. Because BAO observables constrain ratios of the form $D_M(z)/r_d$ and $D_H(z)/r_d$, their dominant sensitivity is to the product $h\,r_d$ rather than to $H_0$ alone. When $r_d$ is allowed to vary freely, variations in $H_0$ can be compensated by shifts in $r_d$, restoring an approximate degeneracy in the $(h,r_d)$ subspace.

To understand how this degeneracy affects the expansion-rate constraint, we project the Fisher tensors onto the physical expansion-rate axis. We perform this projection in two parameter bases, $(\Omo,\ln H_0,w)$ and $(\Omo,H_0, w)$, in order to verify that the resulting geometry is independent of the coordinate representation.

\subsubsection{$\ln H_0$ basis}
\label{subsubsec:621}
\begin{figure}[t]
\centering
\includegraphics[width=0.48\linewidth]
{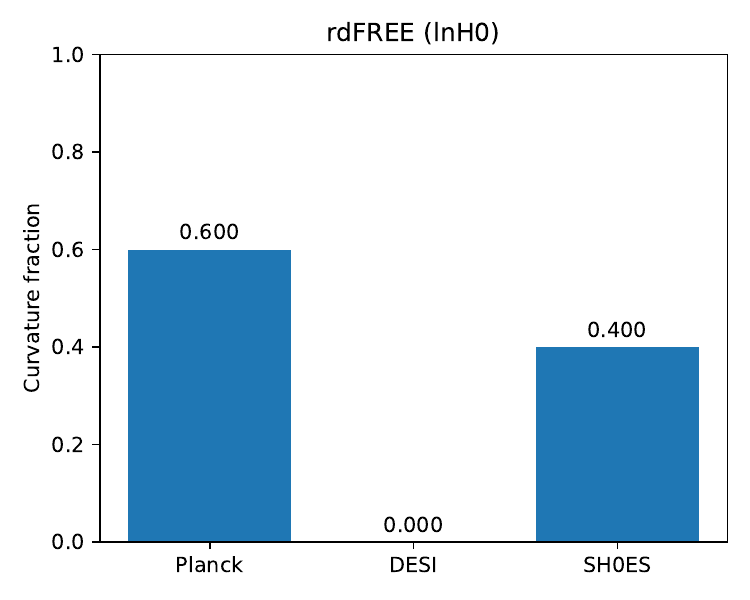}
\hfill
\includegraphics[width=0.48\linewidth]
{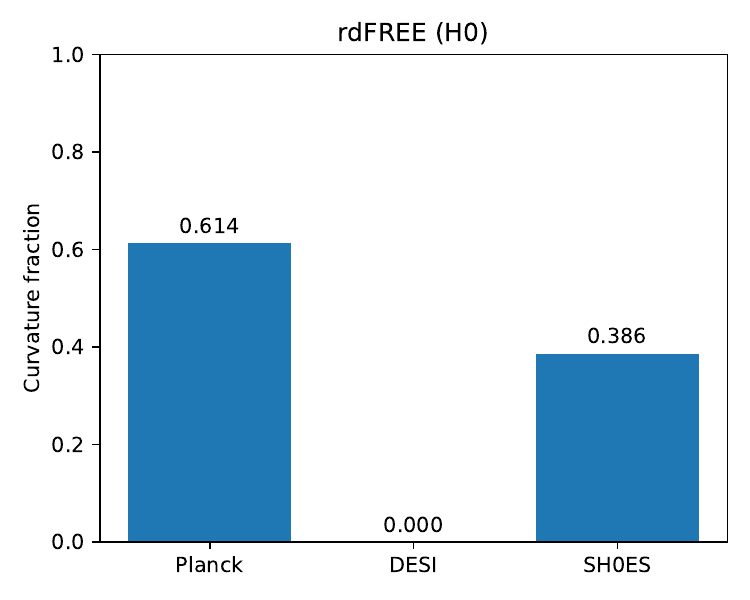}
\caption{Directional curvature fractions in the rdFREE configuration. \textbf{Left}: projection along $\ln H_0$. \textbf{Right}: projection along $H_0$. In both parameter bases DESI contributes zero curvature along the expansion-rate axis. The effective stiffness reduces to a binary Planck-SH0ES balance, with relative contributions of approximately $60{:}40$ in the $\ln H_0$ basis and $61{:}39$ in the $H_0$ basis.}
\label{fig:fraction_rdFREE_combined} 
\end{figure}

In the $(\Omo,\ln H_0,w)$ basis, the directional curvature supplied by probe $X$ is defined as
\begin{equation}
\kappa_X(\ln H_0) = \hat{\mathbf u}_{\ln H_0}^{\mathsf T} \mathbf F_X \hat{\mathbf u}_{\ln H_0},
\end{equation}
where $X=P,D,S$ denote Planck, DESI, and SH0ES. For the rdFREE configuration, we obtain
\begin{align}
\kappa_P = 7342.50,  \qquad \kappa_D = 0, \qquad \kappa_S = 4900.00 .
\end{align}
The corresponding curvature fractions are
\begin{equation}
f_P^{\rm rdFREE} = 0.600, \qquad f_D^{\rm rdFREE} = 0, \qquad f_S^{\rm rdFREE} = 0.400 .
\end{equation}
The resulting curvature partition is shown in the left panel of Fig.~\ref{fig:fraction_rdFREE_combined}. Because $r_d$ is marginalized, DESI contributes no curvature along the pure $\ln H_0$ direction. This vanishing curvature is not a numerical artifact. When $r_d$ is marginalized, BAO observables constrain only the combination $h\,r_d$, leaving the pure $H_0$ direction exactly degenerate in the Fisher projection. Therefore, the joint stiffness is entirely supplied by Planck and SH0ES. The joint directional curvature reduces to
\begin{equation}
\kappa_{\rm joint}^{\rm rdFREE}(\ln H_0) = \kappa_P+\kappa_S , \label{kappajointrdFREE}
\end{equation}
so that the nominal three-probe system becomes geometrically equivalent to a binary curvature balance between the early-Universe acoustic constraint (Planck) and the local distance ladder (SH0ES).

\subsubsection{$H_0$ basis}
\label{subsubsec:622}
We now repeat the projection in the linear $(\Omo,H_0,w)$ basis. The resulting directional curvatures are
\begin{align}
\kappa_P = 1.588, \qquad \kappa_D = 0, \qquad \kappa_S = 1.000 .
\end{align}
The corresponding fractions are
\begin{equation}
f_P^{\rm rdFREE}=0.614, \qquad f_D^{\rm rdFREE}=0, \qquad f_S^{\rm rdFREE}=0.386 .
\end{equation}
These values are shown in the right panel of Fig.~\ref{fig:fraction_rdFREE_combined}. Although the absolute curvatures differ from those in the $\ln H_0$ representation because the two coordinates are related by the nonlinear transformation $dH_0 = H_0\,d\ln H_0$, the qualitative structure is unchanged. DESI remains completely degenerate along the pure expansion-rate axis, and the effective stiffness is again supplied entirely by Planck and SH0ES. The joint curvature is the same as that of Eq.~\eqref{kappajointrdFREE}. This demonstrates that the reduction to an effective two-probe geometry is independent of the chosen parameter basis. The comparison between the $H_0$ and $\ln H_0$ bases serves as a consistency test of the coordinate representation. Although the absolute curvature values change under nonlinear parameter transformations, the physical curvature attribution remains invariant.

\subsection{rdFIX: Restoration of a Genuine Three-Probe Curvature Geometry}
\label{subsec:63}
We now consider the rdFIX configuration, in which the sound horizon $r_d$ is externally calibrated and no longer marginalized in the BAO likelihood. Throughout this analysis, we adopt a fixed value $r_d = 147.05\,\mathrm{Mpc}$, consistent with the DESI DR2 calibration based on the fitting formula of~\cite{Brieden:2022heh}. Because BAO observables depend on the ratios $D_M(z)/r_d$ and $D_H(z)/r_d$, fixing $r_d$ removes the degeneracy between the Hubble parameter and the sound horizon that characterizes the rdFREE case. Consequently, variations in $H_0$ can no longer be absorbed by shifts in $r_d$. DESI therefore acquires intrinsic Fisher curvature along the physical expansion-rate axis, activating a genuine three-probe curvature geometry.

To quantify this effect, we again project the Fisher tensors onto the expansion-rate direction and examine the resulting curvature attribution in both the $(\Omo,\ln H_0,w)$ and $(\Omo,H_0,w)$ parameter bases.

\subsubsection{$\ln H_0$ basis}
In the $(\Omo,\ln H_0,w)$ basis, the projected directional curvatures are
\begin{align}
\kappa_P = 7342.50, \qquad \kappa_D = 107416.91, \qquad \kappa_S = 4900.00 .
\end{align}
The corresponding curvature fractions are
\begin{equation}
f_P^{\rm rdFIX}=0.061, \qquad f_D^{\rm rdFIX}=0.898, \qquad f_S^{\rm rdFIX}=0.041 .
\end{equation}
The curvature partition is shown in the left panel of Fig.~\ref{fig:fraction_rdFIX_combined}. In contrast to the rdFREE configuration, DESI now contributes the dominant share of the expansion-rate stiffness, providing nearly $90\%$ of the total directional curvature. The joint curvature becomes
\begin{equation}
\kappa_{\rm joint}^{\rm rdFIX} = \kappa_P+\kappa_D+\kappa_S . \label{kappardFIX}
\end{equation}
Thus, the expansion-rate constraint is no longer governed by a two-probe balance. Instead, the Fisher geometry is dominated by the BAO contribution supplied by DESI.  This $90\%$ dominance of the directional stiffness by a single probe (DESI) provides the physical mechanism for the high suspiciousness reported in recent combined analyses \cite{Ong:2026tta}. When the joint curvature is so heavily dominated by one probe, any displacement from another probe (like SH0ES) is heavily penalized along that stiff axis, leading to a localized tension signature ($d_G \approx 1$) in parameter space.

\subsubsection{$H_0$ basis}
\begin{figure}[t]
\centering
\includegraphics[width=0.48\linewidth]
{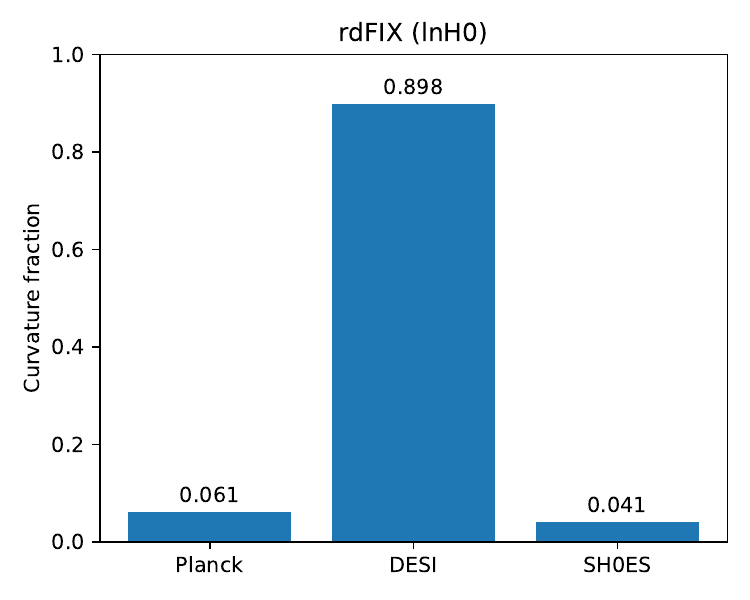}
\hfill
\includegraphics[width=0.48\linewidth]
{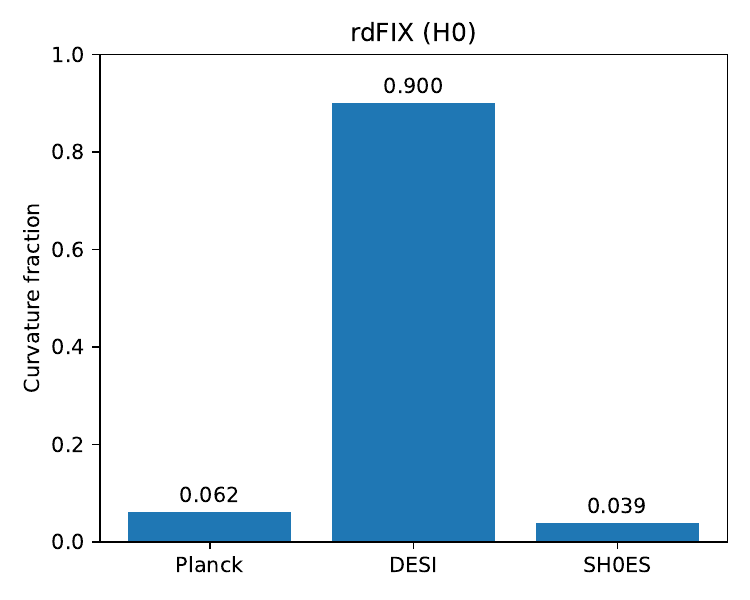}
\caption{Directional curvature fractions in the rdFIX configuration. \textbf{Left}: projection along $\ln H_0$. \textbf{Right}: projection along $H_0$. Fixing the sound horizon breaks the $(h,r_d)$ degeneracy present in rdFREE, allowing DESI to contribute directly to the expansion-rate curvature. In both parameter bases, DESI supplies nearly $90\%$ of the total stiffness along the expansion-rate axis, while Planck and SH0ES provide only minor contributions.}
\label{fig:fraction_rdFIX_combined} 
\end{figure}

We repeat the same projection in the linear $(\Omo,H_0,w)$ basis. The directional curvatures become
\begin{align}
\kappa_P = 1.588, \qquad \kappa_D = 23.228, \qquad \kappa_S = 1.000 .
\end{align}
The resulting curvature fractions are
\begin{equation}
f_P^{\rm rdFIX}=0.062, \qquad f_D^{\rm rdFIX}=0.900, \qquad f_S^{\rm rdFIX}=0.039 .
\end{equation}
These values are shown in the right panel of Fig.~\ref{fig:fraction_rdFIX_combined}. Although the absolute curvature values differ from those in the logarithmic representation, the qualitative structure is unchanged: DESI dominates the directional stiffness along the expansion-rate axis. The joint curvature is also given by Eq.~\eqref{kappardFIX}. This behavior is not specific to the fiducial parameter values, but follows directly from the structure of BAO observables, which constrain the ratios $D_M/r_d$ and $D_H/r_d$.

\subsection{Role of SH0ES: Intrinsic Axial Curvature in Expansion-Rate Space}
\label{subsec:64}
SH0ES differs qualitatively from Planck and DESI in both observational methodology and Fisher-geometry structure. Planck constrains the expansion rate indirectly through preservation of the acoustic angular scale $\theta_\ast$, while DESI constrains mixed parameter combinations involving $(h,r_d)$ through BAO distance ratios. In contrast, SH0ES determines the present-day expansion rate through a locally calibrated distance ladder expressed in absolute physical units. At sufficiently low redshift ($z\ll1$), the luminosity distance satisfies
\begin{equation}
D_L(z) \simeq \frac{c\,z}{H_0},
\end{equation}
where $c$ is the speed of light in vacuum. The distance modulus
\begin{equation}
\mu = 5\log_{10}\!\left(\frac{D_L}{10\,\mathrm{pc}}\right)
\end{equation}
relates the observed apparent magnitude to the absolute luminosity. Cepheid calibration determines the supernova absolute magnitude $M_B$, thereby fixing the normalization of $D_L$ and directly determining $H_0$.

\subsubsection{Fisher--geometric interpretation}
\label{subsubsec:641}
Because $D_L \propto H_0^{-1}$ at low redshift, variations in $H_0$ (or equivalently $\ln H_0$) simply rescale predicted luminosity distances without requiring correlated shifts in other cosmological parameters. Consequently, the SH0ES Fisher tensor projects almost entirely onto the pure expansion-rate axis. In the $(\Omo,\ln H_0,w)$ basis, the directional curvature supplied by SH0ES is
\begin{equation}
\kappa_S(\ln H_0) = \hat{\mathbf u}_{\ln H_0}^{\mathsf T} \mathbf F_S \hat{\mathbf u}_{\ln H_0}.
\end{equation}
Numerically we obtain
\begin{equation}
\kappa_S^{\rm rdFREE} = \kappa_S^{\rm rdFIX} = 4900 ,
\end{equation}
indicating that the SH0ES curvature is unaffected by the treatment of the sound horizon in the BAO likelihood. In the linear $(\Omo,H_0,w)$ basis, the corresponding projection is
\begin{equation}
\kappa_S(H_0) = \hat{\mathbf u}_{H_0}^{\mathsf T} \mathbf F_S \hat{\mathbf u}_{H_0},
\end{equation}
with
\begin{equation}
\kappa_S = 1.0 .
\end{equation}
As expected from the relation $dH_0 = H_0\,d\ln H_0$, the two curvature measures differ only by the coordinate rescaling between the logarithmic and linear representations.

\subsubsection{Numerical construction of the SH0ES Fisher tensor}
\label{subsubsec:642}
The SH0ES Fisher matrix is constructed directly from the Pantheon+SH0ES likelihood using the publicly released distance-modulus data vector and covariance matrix. Unlike the Planck and DESI Fisher tensors, which are reconstructed from posterior covariances of MCMC chains, the SH0ES Fisher tensor is evaluated by numerical differentiation of the supernova distance modulus with respect to the parameters $(H_0,\Omo,M)$. Specifically, the Jacobian matrix $J_{i\alpha}=\partial \mu_i / \partial \theta_\alpha$ is computed using central finite differences, and the Fisher tensor is obtained as
\begin{equation}
F_{\alpha\beta} = J^{\mathsf T} C^{-1} J + F_{\rm prior},
\end{equation}
where $C$ is the full Pantheon+SH0ES covariance matrix and $F_{\rm prior}$ represents the Gaussian calibration prior on the absolute magnitude $M$. The numerical derivatives are evaluated with step sizes $\Delta H_0 = 10^{-3}$, $\Delta\Omo=10^{-4}$, and $\Delta M = 10^{-4}$. Further details of the numerical implementation are summarized in Appendix~\ref{app:numerical_config}.

\subsubsection{Structural role in rdFREE and rdFIX}
\label{subsub643}
The invariance of $\kappa_S$ under the rdFREE/rdFIX transformation highlights the structural distinction of SH0ES within the combined Fisher geometry.

In the rdFREE configuration, DESI contributes no curvature along the expansion-rate axis (Sec.~\ref{subsec:62}), and the effective stiffness is supplied entirely by Planck and SH0ES. The expansion-rate constraint reduces to a binary curvature balance between the early-Universe acoustic scale and the local distance ladder. In the rdFIX configuration, breaking the $(h,r_d)$ degeneracy activates a large DESI contribution to the same axis (Sec.~\ref{subsec:63}). The Fisher geometry then becomes genuinely three-probe, with DESI providing the dominant stiffness while SH0ES remains a geometrically aligned but subdominant contributor.

Thus, SH0ES plays a uniquely stable role within the combined curvature structure. Unlike Planck and DESI, whose projections depend on the acoustic calibration through $r_d$, the SH0ES curvature is intrinsic to the physical expansion-rate axis itself. Its contribution therefore remains unchanged under the rdFREE–rdFIX transformation and remains aligned with the $H_0$ direction in parameter space.

\begin{figure}[t]
\centering
\includegraphics[width=0.75\linewidth]{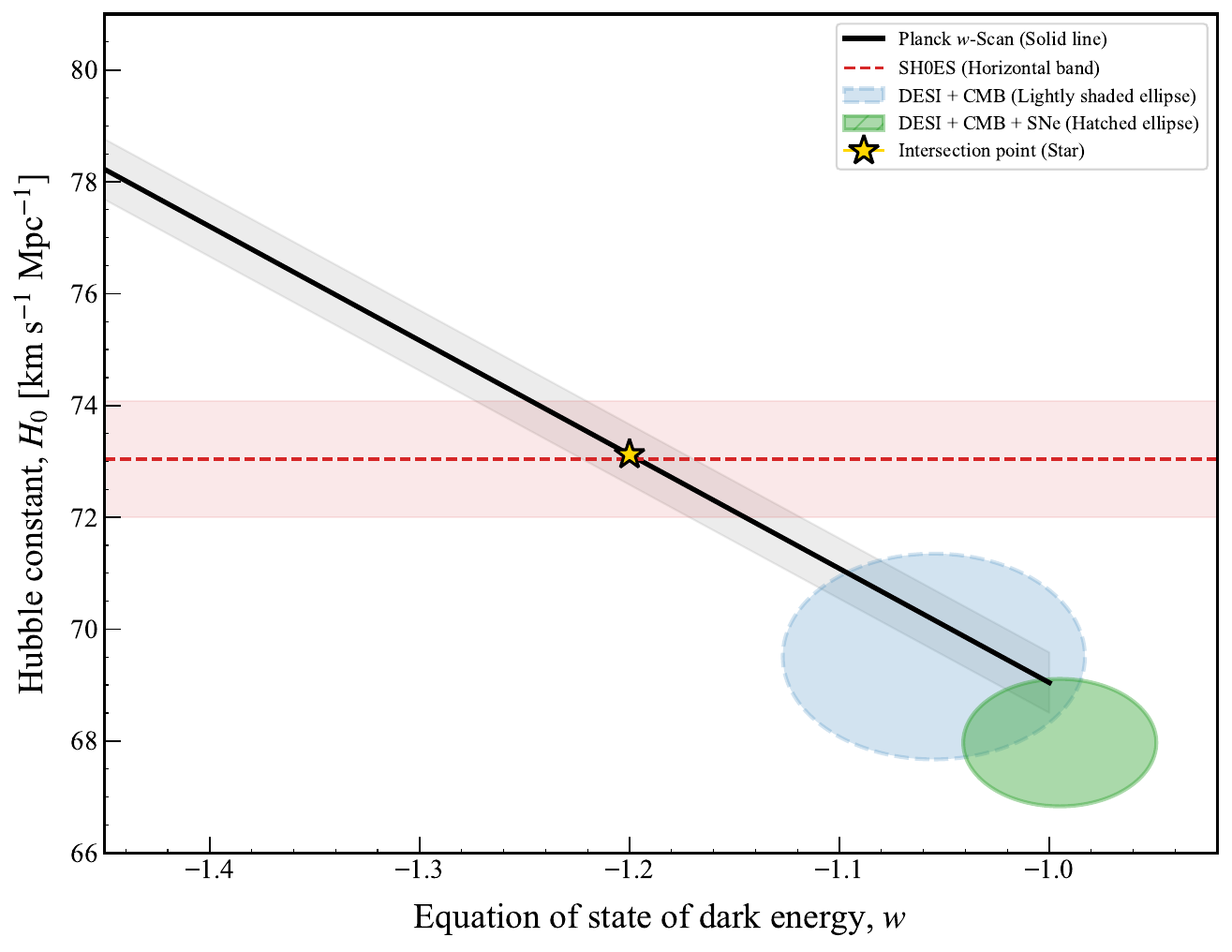}
\caption{Illustrative geometric representation of the three–probe constraint structure in the $(w,H_0)$ plane. The solid curve shows the Planck degeneracy obtained by scanning the dark energy equation of state parameter $w$. The horizontal band indicates the SH0ES measurement of $H_0$, while the ellipses represent the DESI+CMB and DESI+CMB+SNe constraints. The intersection illustrates how late-time datasets relocate the preferred cosmological solution within the Planck degeneracy manifold.}
\label{fig:three_probe_geometry} 
\end{figure}
To illustrate the geometric origin of the multi-probe interaction, Fig.~\ref{fig:three_probe_geometry} presents a schematic representation of the constraint manifolds in the $(w,H_0)$ plane constructed from representative observational results. The solid curve corresponds to a Planck $w$--scan obtained from the \texttt{base\_w} Planck 2018 likelihood (\texttt{TT+low$\ell$+lowE}), which traces the degeneracy direction along which the CMB likelihood preserves the acoustic scale. The horizontal band indicates the SH0ES determination of the Hubble constant, centered on its reported best-fit value. The shaded ellipses represent the DESI DR2 constraints derived from the combinations reported in Table~V of the DESI analysis, including the DESI+CMB and DESI+CMB+SNe datasets.

The figure highlights a key geometric feature of the parameter space. The Planck degeneracy direction favors a displacement toward the phantom regime ($w<-1$) when attempting to reconcile the CMB solution with the SH0ES measurement of $H_0$. In contrast, the DESI constraints are centered on the quintessence side ($w>-1$), with the DESI-only analysis yielding
$w=-0.916\pm0.078$. Consequently, the preferred directions of the Planck and DESI curvature contributions are nearly opposite in the $(w,H_0)$ plane. When these constraints are combined, the competing curvature gradients intersect near $w\simeq -1$, producing a geometrically stabilized solution close to $\Lambda$CDM. In Fisher-geometric terms, this behavior corresponds to a curvature collision,  in which the directional stiffness supplied by DESI counteracts the phantomward degeneracy of the Planck likelihood. The figure provides an intuitive visualization of the curvature redistribution discussed above and motivates the unified geometric interpretation presented in the following subsection. This curvature collision explains why the frequentist preference for $w < -1$ disappears in a full Bayesian treatment \cite{Ong:2026tta}. While the Planck degeneracy pulls toward the phantom regime, the stiff DESI curvature acts as a restoring force toward $\Lambda$CDM. The intersection near $w \simeq -1$ is not a coincidence but a geometric stabilization where competing curvature gradients reach a stalemate, favoring the simpler $\Lambda$CDM model via Ockham's razor.

\subsection{Unified geometric interpretation}
\label{subsec:65}
We may now synthesize the preceding results into a unified geometric picture. Throughout this analysis, distances are expressed in absolute physical units, and the Hubble tension corresponds to a displacement $\Delta H_0$ (or equivalently $\Delta\ln H_0$) along a physically meaningful expansion-rate axis. Its statistical weight is governed by the quadratic form
\begin{equation}
T^2(\hat{\mathbf u}) = \kappa_{\rm joint}(\hat{\mathbf u}) \, (\Delta\boldsymbol{\theta}\cdot\hat{\mathbf u})^2 \,,
\end{equation}
so that the significance of the tension is controlled not only by the magnitude of the parameter shift but also by the directional stiffness that weights that shift.

\begin{figure}[t]
\centering
\includegraphics[width=0.48\linewidth]
{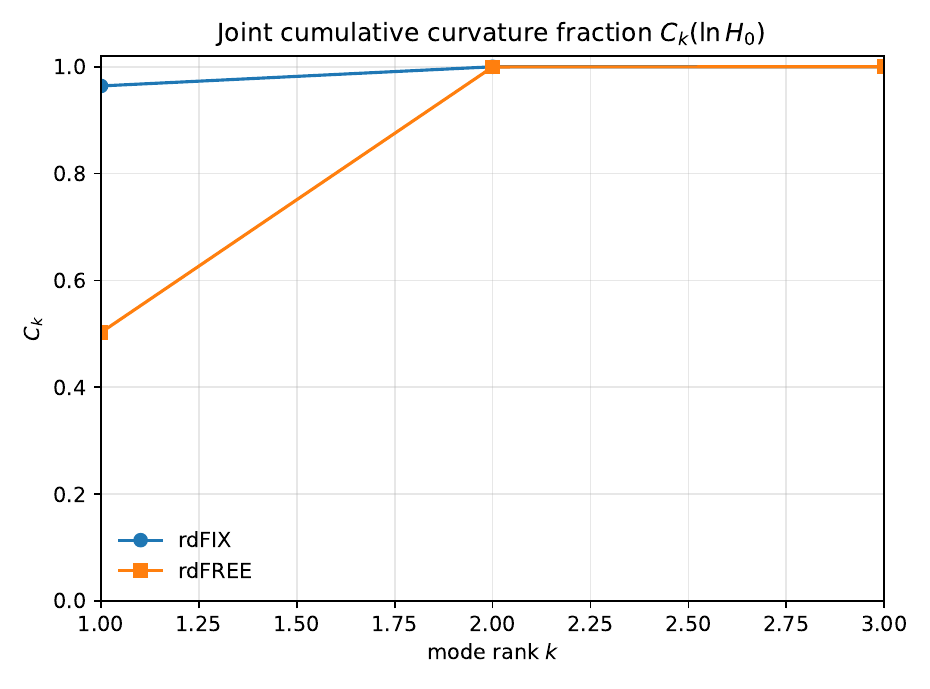}
\hfill
\includegraphics[width=0.48\linewidth]
{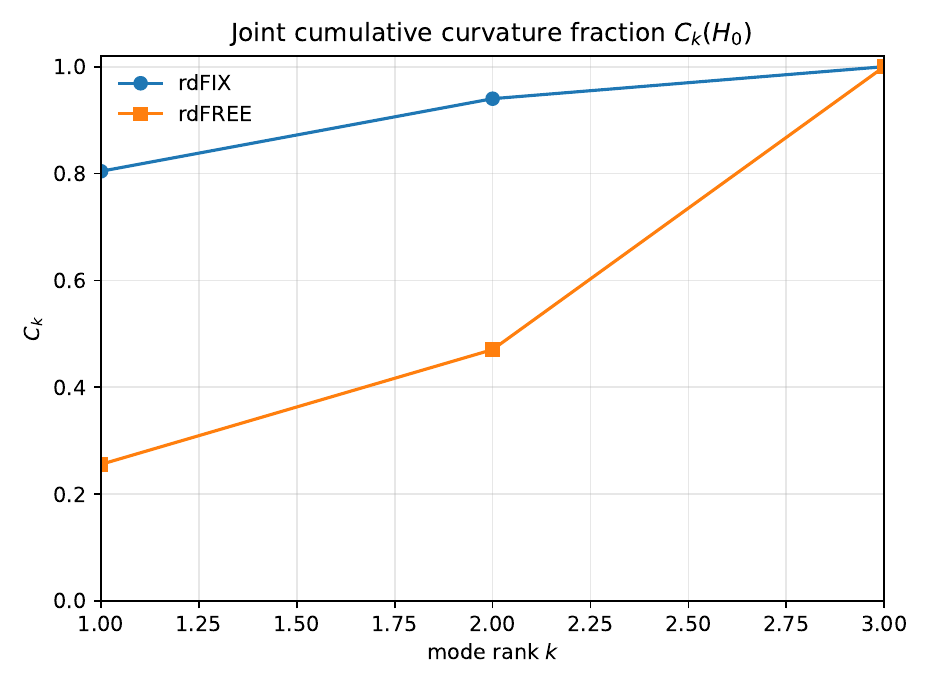}
\caption{Cumulative curvature fraction $C_k$ of the joint Fisher tensor. \textbf{Left}: $C_k(\ln H_0)$. \textbf{Right}: $C_k(H_0)$. The saturation scale of the cumulative curvature depends on both the parameter representation and the treatment of the sound horizon. For $\ln H_0$, the curvature is captured by the first mode in the rdFIX configuration and by the first two modes in rdFREE. In the linear $H_0$ basis, the curvature is distributed across several leading modes, approaching saturation only after the first few eigenmodes of the Fisher spectrum.}
\label{fig:Ck_rd_combined} Fig6
\end{figure}

\subsubsection{Low-dimensional curvature structure}
\label{subsubsec:651}
Figure~\ref{fig:Ck_rd_combined} shows the cumulative curvature fraction $C_k$ of the joint Fisher tensor. In the $\ln H_0$ representation, the curvature saturates rapidly, reaching unity by the first mode in rdFIX and by the second mode in rdFREE. In the linear $H_0$ basis, the curvature is distributed over a slightly larger set of leading modes, approaching saturation only after the first few eigenmodes of the Fisher spectrum.

Even under model extension and under rdFREE/rdFIX transformations, the expansion-rate stiffness therefore remains confined to a very low-dimensional subspace of parameter space. Although more than a single eigenmode contributes in some parameter representations, the curvature is still dominated by only a small number of leading Fisher modes.

Although the orientation of the stiff direction may change, the hierarchical structure of the eigenvalue spectrum remains intact. The moderate eigenvector rotations between rdFREE and rdFIX,
\begin{align}
\Delta\theta_{\rm lead}(\ln H_0) \simeq 24.0^\circ, \qquad \Delta\theta_{\rm lead}(H_0) \simeq 10.7^\circ ,
\end{align}
modify the orientation of the dominant constraint, but do not significantly alter the curvature hierarchy. Therefore, the dominant stiffness scale remains concentrated in the leading Fisher modes.

\subsubsection{Redistribution of stiffness}
\label{subsubsec:652}
Within this anisotropic Fisher geometry, the treatment of the sound horizon controls how stiffness is redistributed among probes. In the rdFREE configuration, DESI contributes identically zero curvature along the pure expansion-rate axis, and the joint curvature reduces to $\kappa_{\rm joint}^{\rm rdFREE} = \kappa_P + \kappa_S$. Numerically, the stiffness partitions approximately as
$(f_P,f_D,f_S) = (0.60,\,0,\,0.40)$ for $\ln H_0$, and $(0.61,\,0,\,0.39)$ for $H_0$. Therefore, the geometry reduces to an effective binary balance between early-Universe acoustic rigidity (Planck) and local distance-ladder stiffness (SH0ES).

When the sound horizon is externally calibrated (rdFIX), the $(h,r_d)$ degeneracy is broken and DESI acquires a finite projection onto the expansion-rate axis. The joint curvature then becomes $\kappa_{\rm joint}^{\rm rdFIX} = \kappa_P + \kappa_D + \kappa_S$  with stiffness fractions $(f_P,f_D,f_S) = (0.06,\,0.90,\,0.04)$ for both $\ln H_0$ and $H_0$. DESI becomes the dominant source of expansion-rate rigidity, and the system realizes a genuine three-probe curvature geometry.

\subsubsection{Geometric origin of the tension}
\label{subsubsec:653}
The essential insight is geometric. The Hubble tension is not merely a discrepancy between two central values of $H_0$; it is the projection of a physical displacement onto a highly anisotropic Fisher metric whose stiffness is concentrated in a single eigenmode. Sound-horizon calibration determines whether that stiffness is shared primarily between Planck and SH0ES (rdFREE) or redistributed to include a dominant DESI contribution (rdFIX).  Thus, the character of the tension — binary versus genuinely three-probe —is controlled not by the numerical value of $H_0$ alone, but by how Fisher curvature is partitioned within a low-dimensional information geometry of the parameter space.  Our geometric diagnostics allow for a calculation-free understanding of the Hubble tension: by identifying the partition of Fisher curvature ($f_X$), one can predict the stability of the joint solution without exhaustive MCMC sampling. The shift from a binary Planck--SH0ES balance to a DESI-dominated three-probe geometry is the fundamental reason why recent DESI DR2 results have stabilized the $\Lambda$CDM framework against various model extensions.

\subsubsection{Connection to observational constraints}
\label{subsubsec:654}
This geometric structure is reflected in current cosmological parameter constraints.  Although allowing the dark energy equation of state to vary slightly shifts the preferred expansion rate, the combined DESI+CMB+Pantheon+ constraints in the $w$CDM model yield~\cite{DESI:2025zgx}
\begin{equation}
H_0 = 67.97 \pm 0.57 \,\mathrm{km\,s^{-1}\,Mpc^{-1}} .
\end{equation}
This value remains substantially below the local distance-ladder determination by SH0ES, $H_0 = 73.04 \pm 1.04 \,\mathrm{km\,s^{-1}\,Mpc^{-1}}$, corresponding to a discrepancy of approximately $4.3\sigma$. Within the geometric framework developed here, this stability arises from the strong directional curvature associated with the dominant Fisher eigenmode, which restricts the displacement of the expansion-rate parameter even under modest extensions of the cosmological model.

\begin{table*}[t]
\centering
\caption{Structural diagnostics for the $\ln H_0$ direction in the unified $(\Omo,\ln H_0,w)$ basis. The table summarizes how each probe contributes to the directional curvature along the expansion-rate axis under the rdFREE and rdFIX configurations.}
\label{tab:H0_diagnostics}
\begin{tabular}{lcccc}
\hline\hline
Dataset & Alignment & Localization & $C_k$ Structure & Geometric Role \\
\hline
Planck & Leading mode & Hierarchical & Leading-mode saturation & Acoustic rigidity \\
SH0ES & Leading mode & Axial concentration & Leading-mode saturation & Axial stiffness \\
DESI (rdFREE) & No projection & Degenerate in $H_0$ & No contribution & Inactive \\
DESI (rdFIX) & Leading mode & Single stiff mode & Leading-mode saturation & Dominant constraint
\\
\hline\hline
\end{tabular}
\end{table*}
Table~\ref{tab:H0_diagnostics} summarizes the structural distribution of curvature across eigenmodes, highlighting the low-dimensional character of the expansion-rate geometry and its redistribution under sound-horizon calibration.

\subsection{Summary of Curvature Diagnostics}
\label{subsec:66}
The directional curvature contributions along the expansion-rate axis are summarized in Table~\ref{tab:curvature_summary}. All values correspond to the Jacobian-consistent projections reported in Sec.~\ref{subsec:61}. The table compares the curvature amplitudes in both the $\ln H_0$ and $H_0$ parameter bases. Although the absolute numerical values differ because of the
nonlinear coordinate transformation $dH_0 = H_0\,d\ln H_0$, the geometric structure and curvature partitioning remain unchanged.

In the rdFREE configuration, the system reduces to a binary Planck–SH0ES curvature balance, with DESI contributing no stiffness along the pure expansion-rate axis. When the sound horizon is externally calibrated (rdFIX), the $(h,r_d)$ degeneracy is broken, and DESI becomes the dominant source of directional curvature. Therefore, the Hubble tension reflects the redistribution of curvature within a highly anisotropic Fisher geometry rather than a discrepancy in central parameter values alone.

\begin{table*}[t]
\centering
\caption{Directional curvature diagnostics along the expansion-rate axis. Curvatures are shown for both the logarithmic ($\ln H_0$) and linear ($H_0$) parameter bases. The two are related by the Jacobian transformation $\kappa(\ln H_0)=H_0^2\,\kappa(H_0)$ evaluated at the fiducial $H_0$. Fractions $f_X$ refer to the contribution of each probe to the total directional curvature.}
\label{tab:curvature_summary}
\begin{tabular}{lcccc}
\hline\hline
Probe & $\kappa(\ln H_0)$ & $\kappa(H_0)$ & Fraction $f_X$ & Geometric Role \\
\hline
\multicolumn{5}{c}{\textit{rdFREE}} \\[2pt]
Planck & $7342.5$ & $1.588$ & $0.60$ & Acoustic rigidity \\
SH0ES & $4900.0$ & $1.000$ & $0.40$ & Axial stiffness \\
DESI & $0$ & $0$ & $0$ & Degenerate along $H_0$ \\
\hline
\multicolumn{5}{c}{\textit{rdFIX}} \\[2pt]
Planck & $7342.5$ & $1.588$ & $0.06$ & Subdominant acoustic contribution \\
SH0ES & $4900.0$ & $1.000$ & $0.04$ & Minor axial contribution \\
DESI & $107416.9$ & $23.228$ & $0.90$ & Dominant geometric constraint \\
\hline\hline
\end{tabular}
\end{table*}

\section{Discussion: Geometric Conditions for Alleviating the $H_0$ Tension}
\label{sec:discussion}

The preceding sections established a quantitative decomposition of the $H_0$ tension within a unified Fisher--geometric framework. We showed that the statistical weight of the expansion-rate direction is highly anisotropic and confined to a very low-dimensional subspace of parameter space, with the cumulative curvature saturating after only the first few leading Fisher modes.

Rather than interpreting the tension simply as a discrepancy between two central values of $H_0$, the Fisher--geometric framework identifies it as the projection of a parameter displacement onto an anisotropic information metric. In this view, the statistical significance of the tension is governed jointly by the magnitude of the parameter displacement and by the directional stiffness supplied by the data. This section discusses the physical implications of this geometric picture, identifies the conditions under which the tension can be reduced, and highlights the convergence between our geometric diagnostics and recent Bayesian evidence in the context of our prior systematic studies~\cite{Lee:2025jrr,Lee:2025kbn,Lee:2025rmg,Lee:2025ysg,Lee:2025grb,Lee:2025axp,Lee:2026sta}.

\subsection{The Geometric Nature of the $H_0$ Tension}
\label{subsec:71}

Let $\boldsymbol{\theta}$ denote the cosmological parameter vector and $\hat{\mathbf u}$ a unit direction in parameter space. In the Gaussian approximation, the statistical significance of a projected displacement between probes is given by $T^2(\hat{\mathbf u}) = \kappa_{\rm joint}(\hat{\mathbf u}) \big(\Delta\boldsymbol{\theta}\cdot\hat{\mathbf u}\big)^2$, where $\kappa_{\rm joint}$ is the directional curvature supplied by the combined datasets. This factorization clarifies that even a small displacement can become statistically significant if it lies along a stiff axis of the Fisher metric.

Our analysis in Sec.~\ref{sec:three_probe} demonstrated that the expansion-rate direction is controlled by a strongly hierarchical eigenvalue spectrum. This low-dimensional structure is consistent with our redshift-resolved diagnostics~\cite{Lee:2025rmg}, which showed that individual DESI DR2 BAO bins remain remarkably consistent with $\Lambda$CDM. This geometric rigidity is further reflected in the low Bayesian model dimensionality ($d_G \approx 1.5 - 3.5$) reported in recent nested sampling analyses~\cite{Ong:2026tta}. These synergistic results confirm that the $H_0$ discordance is not a systemic multi-dimensional discrepancy but a localized conflict along a specific high-stiffness axis of the information metric.

\subsection{Why Enlarging Parameter Space Often Fails}
\label{subsec:72}

Model extensions modify cosmological inference by altering the Fisher curvature tensor. However, as rigorously derived in our linear-response approach~\cite{Lee:2025ysg}, the sensitivity of distance observables to the time evolution of dark energy is intrinsically limited by the integral structure of the Hubble rate. This structural limitation explains why introducing additional parameters like $w_a$ often fails to "soften" the dominant stiffness scale effectively.

While a shift toward a phantom regime ($w \approx -1.2$) can mathematically reconcile Planck with SH0ES ($T^2 \approx 0.07\sigma$), our analysis of the CPL parametrization~\cite{Lee:2025ysg} indicates that such escape routes are highly sensitive to the choice of data basis and prior assumptions. If the leading eigenvalue remains large, the rigidity along the principal direction is preserved. This persistence is evident in combined DESI+CMB+Pantheon+ analyses~\cite{DESI:2025zgx}, where the joint solution is forced back toward $\Lambda$CDM-like values by the combined curvature of the probes, acting as a geometric wall against large phantom shifts.

\subsection{Synergy with Bayesian Evidence and Systematics}
\label{subsec:73}

Recent Bayesian reanalyses~\cite{Ong:2026tta,Ormondroyd:2025phk,Ong:2025utx} concluded that the preference for dynamical dark energy (DDE) vanishes when proper model selection and calibration uncertainties are accounted for ($\ln B \approx -0.6$). These findings provide an independent statistical validation of our earlier diagnosis regarding the impact of $\Omo$ prior bias~\cite{Lee:2025kbn}. We demonstrated that imposing informative $\Omo$ priors from supernova data can artificially shift the equation-of-state parameters away from $\Lambda$CDM—a result of the inherent projection effects within a highly anisotropic likelihood geometry. 

Furthermore, our pedagogical null tests~\cite{Lee:2025grb} and tension forecasts~\cite{Lee:2025axp} showed that residual inter-probe inconsistencies can mimic evidence for DDE. The recently identified DES-SN5YR calibration error~\cite{Ong:2026tta,Ormondroyd:2025phk} is therefore fully consistent with our framework, in which extended parameter directions absorb cross-probe inconsistencies. 

It is useful to distinguish our geometric analysis from recent Bayesian discussions by Cort\^{e}s and Liddle~\cite{Cortes:2023dij,Cortes:2024lgw,Cortes:2025joz}. While they emphasize that reduced tension alone does not support new physics without accounting for Bayesian evidence, our work addresses the complementary question: it identifies the internal geometric mechanism that produces or suppresses the tension. The Bayesian perspective clarifies when reduced tension should be trusted, while the Fisher--geometric framework explains why the curvature structure produces the tension in the first place.

\subsection{Sound-Horizon Calibration as a Geometric Test}
\label{subsec:74}

The treatment of the sound horizon ($r_d$) provides a clear illustration of curvature redistribution. In the rdFREE configuration, the stiffness is supplied by a balance between Planck and SH0ES. In the rdFIX configuration, the BAO likelihood acquires a significant projection onto the expansion-rate axis, making DESI the dominant source of rigidity.

As shown in~\cite{Lee:2025rmg}, this anchoring of $r_d$ is crucial for the robustness of $\Lambda$CDM. The negative Bayesian suspiciousness ($\log S \approx -2.7$) reported in recent combined analyses~\cite{Ong:2026tta} during this transition further underscores that the tension is governed by the curvature partition among probes. This redistribution of stiffness, rather than a large multi-dimensional discrepancy, characterizes the transition from a binary to a genuinely three-probe geometry.

\subsection{Physical Pathways Toward Tension Reduction}
\label{subsec:75}

Within the geometric framework, alleviating the $H_0$ tension requires modifying either the projected displacement ($\Delta$) or the stiffness ($\kappa$) that weights it. We identify three broad classes of mechanisms, summarized in Table~\ref{tab:tension_reduction}:
\begin{enumerate}
    \item \textbf{Shift reduction}: Reducing the projected displacement along the dominant stiffness direction (e.g., sound-horizon modifications like EDE).
    \item \textbf{Curvature softening}: Reducing the leading eigenvalue ($\lambda_1$) of the principal eigenmode.
    \item \textbf{Curvature rebalancing}: Redistributing stiffness among probes to change the weighting of the displacement.
\end{enumerate}

Additionally, our analysis of growth probes~\cite{Lee:2026sta} suggests that while distance measurements provide the primary constraint, growth measurements offer a complementary pathway to break geometric degeneracies, potentially identifying whether shifts in $H_0$ are physical or systematic.

\begin{table*}[t]
\centering
\caption{Geometric pathways capable of alleviating the $H_0$ tension within the shift--stiffness framework developed in this work. The conditions reflect the Fisher eigenvalue hierarchy and curvature-partition results derived in Sec.~\ref{sec:three_probe}, including the rdFREE and rdFIX configurations. The table is intended as a structural diagnostic linking physical scenarios to the geometric conditions required for reducing the statistical significance of the tension.}
\label{tab:tension_reduction}
\begin{tabular}{p{2.6cm} p{3.0cm} p{3.8cm} p{3.4cm} p{2.6cm}} 
\hline\hline
\textbf{Mechanism} & \textbf{Mathematical Condition} & \textbf{Geometric Interpretation} & \textbf{Required Probe / Model Property} & \textbf{Representative Examples} \\
\hline
\textbf{(I) Shift reduction} & $|\Delta\ln H_0| \downarrow$ &
Preferred expansion-rate values move closer
along the dominant stiffness direction.
Posterior broadening without relocation
does not reduce the tension.
&
Model or calibration must shift
the preferred expansion rate
along the principal Fisher eigenmode.
Degeneracy enlargement alone is insufficient.
&
Early-time sound-horizon modifications
(EDE-like models),
improved distance-ladder calibration
\\[1.2em]

\textbf{(II) Curvature softening} & $\lambda_1 \downarrow$
or 
$\kappa_{\rm joint}(H_0)\downarrow$
&
The principal rigidity scale controlling
the expansion-rate direction is reduced.
Broadening orthogonal directions
does not significantly affect the tension.
&
Model must weaken the acoustic-scale
stiffness while remaining consistent with
BAO, lensing, and growth constraints.
The dominant eigenmode must soften.
&
Extended $w$CDM-like models,
modified recombination physics
\\[1.2em]

\textbf{(III) Curvature rebalancing} &
Redistribution of $\kappa_X(H_0)$
(e.g.\ $f_P\downarrow$, $f_L\uparrow$)
&
Additional probes inject curvature
directly along the expansion-rate axis,
changing the stiffness partition
among datasets.
Rebalancing alone does not guarantee
tension reduction if the eigenvalue
hierarchy persists.
&
Observable must have strong
$\partial O/\partial H_0$
and favorable alignment with
the dominant Planck mode.
Injected curvature must compete
with early-Universe stiffness.
&
Standard sirens (GW),
time-delay cosmography,
BAO with fixed $r_d$ (rdFIX)
\\[1.2em]

\textbf{(IV) Degeneracy intersection}
&
Eigenmode alignment between
Planck and late-time probes
(e.g.\ finite rotation angles)
&
Intersection of stiff directions
relocates the joint solution
within a low-dimensional parameter
manifold.
Small rotations have limited impact
when eigenvalue hierarchy is strong.
&
Late-time probe must introduce
a stiff mode intersecting the
Planck degeneracy manifold.
The rotation must be comparable
to the eigenvalue separation scale.
&
Combined geometry–growth probes,
high-precision structure growth
measurements
\\

\hline\hline
\end{tabular}
\end{table*}

\subsection{Concluding Perspective}
\label{subsec:76}

The analysis developed in this work reframes the $H_0$ tension as a geometric property of parameter inference. The constraint is confined to a very low-dimensional subspace of parameter space, with curvature dominated by only the first few Fisher eigenmodes. 

The robustness of $\Lambda$CDM reported in the latest Bayesian studies~\cite{Ong:2026tta} provides a timely corroboration of the stability and prior-bias mechanisms established in our series of works~\cite{Lee:2025jrr,Lee:2025kbn,Lee:2025rmg,Lee:2025ysg,Lee:2025grb,Lee:2025axp,Lee:2026sta}. Our Fisher--geometric framework offers a physical foundation for these findings, explaining why DDE hints often vanish when the dominant curvature mode is properly accounted for. Understanding this metric structure is essential for distinguishing between dataset-level systematics and genuine evidence for new physics.

\section{Conclusion}
\label{sec:conclusion}

In this work, we have examined the $H_0$ tension from the perspective of information geometry. Rather than characterizing the discrepancy solely through posterior widths or differences of best-fit values, we expressed it in terms of two geometric ingredients: a displacement between preferred parameter values (shift), and a directional rigidity encoded in the Fisher curvature (stiffness).

In the Gaussian approximation, the statistical significance of a projected displacement along a direction $\hat{\mathbf u}$ can be written as $T^2(\hat{\mathbf u}) = \kappa_{\rm joint}(\hat{\mathbf u}) (\Delta\boldsymbol{\theta}\cdot\hat{\mathbf u})^2$, where $\kappa_{\rm joint}$ denotes the directional curvature supplied by the combined probes. This decomposition provides a physically transparent and computationally efficient diagnostic, allowing us to identify the origin of the tension directly from the hierarchical structure of the information metric.

Applying this framework to $\Lambda$CDM and $w$CDM leads to three main conclusions:

First, extending $\Lambda$CDM to $w$CDM primarily reshapes the Fisher geometry through curvature suppression. We quantified a dramatic reduction in the leading Planck eigenvalue (by a factor of $\sim 37.5$ in the $(\Omo, \ln H_0, w)$ space), providing a physical explanation for the Ockham penalty observed in recent Bayesian model-selection studies. Since the orientation of the leading eigenmode remains relatively stable, the $w$CDM extension weakens the constraint's rigidity without providing a genuine physical escape from the acoustic-scale manifold.

Second, the curvature budget along the expansion-rate axis is governed by a low-dimensional hierarchy. Our $C_k$ analysis confirms that the directional stiffness is concentrated in a very small number of Fisher eigenmodes. In the rdFIX configuration, DESI DR2 injects nearly $90\%$ of the total directional curvature, effectively establishing a new "geometric wall" that reinforces the $\Lambda$CDM solution. The transition from a binary Planck--SH0ES balance in rdFREE to a DESI-dominated three-probe geometry in rdFIX is a purely geometric redistribution of information that determines the statistical weight of the tension.

Third, the persistence of the $H_0$ tension is a manifestation of geometric collision. The reported robustness of $\Lambda$CDM in the latest observational analyses is a predictable consequence of the competing curvature gradients identified in this work. Absent a genuine reduction in parameter displacement or a significant softening of the principal stiffness scale, additional datasets primarily serve to tighten the existing geometric barriers rather than alleviating the underlying discordance.

Ultimately, our information--geometric framework offers a calculation-free understanding of why cosmological tensions persist or vanish. By identifying that the current Hubble tension is dominated by a single, highly rigid Fisher eigenmode, we have shown that the apparent hints of new physics in extended models are often artifacts of geometric projection and prior-induced shifts.

Future observational programs—such as standard sirens or time-delay cosmography—will be essential to provide independent curvature that can genuinely compete with the established early-Universe hierarchy. Until then, distinguishing between dataset-level systematics and evidence for new physics will benefit from the unified metric understanding of cosmological parameter space developed in this work.

\appendix
\section{Completing the square for an early--late multi-probe Gaussian product}
\label{app:complete_square_multprobe}
In this appendix we present the full algebraic derivation showing how the product of three statistically independent Gaussian likelihoods (Planck, SH0ES, and DESI) can be regrouped into an early-late block form. Specifically, we derive

(i) the late-block curvature
\(
F_{\rm L}=F_{\rm S}+F_{\rm D}
\)
and mean
\(
\theta_{\rm L}=F_{\rm L}^{-1}(F_{\rm S}\theta_{\rm S}+F_{\rm D}\theta_{\rm D})
\),
(ii) the joint curvature
\(
F_{\rm joint}=F_{\rm P}+F_{\rm L}=F_{\rm P}+F_{\rm S}+F_{\rm D}
\)
and mean
\(
\theta_{\rm joint}
=
F_{\rm joint}^{-1}
(F_{\rm P}\theta_{\rm P}+F_{\rm S}\theta_{\rm S}+F_{\rm D}\theta_{\rm D})
\),
and (iii) the early--late quadratic tension
\begin{equation}
T_{\rm joint}^2 = (\theta_{\rm P}-\theta_{\rm L})^{T} \big[ F_{\rm P} (F_{\rm P}+F_{\rm L})^{-1} F_{\rm L} \big] (\theta_{\rm P}-\theta_{\rm L}),
\end{equation}
which represents the exact residual shift penalty at the joint maximum-likelihood point.

Throughout this appendix we adopt the local Gaussian approximation to each likelihood around its maximum, so that the Fisher matrix provides a quadratic expansion of the log-likelihood. While extended models such as $w$CDM may exhibit non-Gaussian degeneracy structures in certain directions, the dominant stiffness direction identified in this work corresponds to the principal Fisher eigenmode. In the vicinity of the maximum-likelihood region, the projected contours in the $(\Omo,H_0)$ subspace are well described by the Fisher ellipses. The geometric interpretation derived from the leading curvature mode remains robust.

\subsection{Product of three Gaussian likelihoods}

Before forming the Gaussian product, all probes are first expressed in a common parameter basis $\boldsymbol{\theta}$ through the appropriate Jacobian transformations, as described in Sec.~\ref{sec:three_probe}.   Only after this basis unification do we construct the Gaussian product.

Consider three Gaussian likelihoods for a common parameter vector $\theta$,
\begin{align}
L_{\rm P}(\theta) &\propto \exp\!\left[ -\frac12(\theta-\theta_{\rm P})^{T} F_{\rm P} (\theta-\theta_{\rm P}) \right], \\
L_{\rm S}(\theta) &\propto \exp\!\left[ -\frac12(\theta-\theta_{\rm S})^{T} F_{\rm S} (\theta-\theta_{\rm S}) \right], \\
L_{\rm D}(\theta) &\propto \exp\!\left[ -\frac12(\theta-\theta_{\rm D})^{T} F_{\rm D} (\theta-\theta_{\rm D}) \right],
\end{align}
where each $F_i$ is symmetric positive-definite. Their product defines the joint likelihood
\begin{equation}
L_{\rm joint}(\theta) \equiv L_{\rm P}(\theta)L_{\rm S}(\theta)L_{\rm D}(\theta) \propto \exp[-\tfrac12 Q_3(\theta)],
\end{equation}
with
\begin{equation}
Q_3(\theta)= (\theta-\theta_{\rm P})^{T}F_{\rm P}(\theta-\theta_{\rm P}) + (\theta-\theta_{\rm S})^{T}F_{\rm S}(\theta-\theta_{\rm S}) + (\theta-\theta_{\rm D})^{T}F_{\rm D}(\theta-\theta_{\rm D}).
\end{equation}

\subsection{Explicit expansion of the exponent}

Using the symmetry of $F_i$ we obtain
\begin{align}
(\theta-\theta_i)^T F_i (\theta-\theta_i) &=
\theta^T F_i \theta
-2\theta^T F_i \theta_i
+\theta_i^T F_i \theta_i .
\end{align}
Summing over the three probes yields

\begin{align}
Q_3(\theta) &=
\theta^{T}(F_{\rm P}+F_{\rm S}+F_{\rm D})\theta
-2\theta^{T}(F_{\rm P}\theta_{\rm P}
+F_{\rm S}\theta_{\rm S}
+F_{\rm D}\theta_{\rm D})
\nonumber\\
&\qquad
+\theta_{\rm P}^{T}F_{\rm P}\theta_{\rm P}
+\theta_{\rm S}^{T}F_{\rm S}\theta_{\rm S}
+\theta_{\rm D}^{T}F_{\rm D}\theta_{\rm D}.
\end{align}

\subsection{Completing the square}

Define
\begin{equation}
A_3 = F_{\rm P}+F_{\rm S}+F_{\rm D},\qquad
b_3 = F_{\rm P}\theta_{\rm P}+F_{\rm S}\theta_{\rm S}+F_{\rm D}\theta_{\rm D}.
\end{equation}
Then
\begin{equation}
Q_3(\theta) = \theta^T A_3 \theta -2\theta^T b_3 +c_3 .
\end{equation}
Completing the square yields
\begin{equation}
Q_3(\theta) = (\theta-\mu_3)^T A_3 (\theta-\mu_3) + (c_3-\mu_3^T A_3 \mu_3)
\end{equation}
with
\begin{equation}
\mu_3 = A_3^{-1} b_3 .
\end{equation}

Thus, the joint likelihood becomes Gaussian with
\begin{equation}
F_{\rm joint}=F_{\rm P}+F_{\rm S}+F_{\rm D},
\end{equation}
and
\begin{equation}
\theta_{\rm joint} = (F_{\rm P}+F_{\rm S}+F_{\rm D})^{-1} (F_{\rm P}\theta_{\rm P}+F_{\rm S}\theta_{\rm S}+F_{\rm D}\theta_{\rm D}).
\end{equation}

\subsection{Early--late regrouping}
Combining the two late-time probes first,
\begin{equation}
F_{\rm L}=F_{\rm S}+F_{\rm D},  \qquad \theta_{\rm L} = (F_{\rm S}+F_{\rm D})^{-1} (F_{\rm S}\theta_{\rm S}+F_{\rm D}\theta_{\rm D}).
\end{equation}
Combining this block with the early probe gives
\begin{equation}
F_{\rm joint}=F_{\rm P}+F_{\rm L},
\end{equation}
which reproduces the full three-probe result.

\subsection{Early--late quadratic tension}
The statistical tension between the early-Universe block (Planck) and the late-Universe block (SH0ES + DESI) is quantified by the residual of the combined quadratic forms evaluated at the joint minimum, $\theta_{\rm joint}$. Defining the individual penalties as $Q_{\rm P}(\theta) = (\theta-\theta_{\rm P})^T F_{\rm P} (\theta-\theta_{\rm P})$ and $Q_{\rm L}(\theta) = (\theta-\theta_{\rm L})^T F_{\rm L} (\theta-\theta_{\rm L})$, the exact tension statistic is
\begin{equation}
T_{\rm joint}^2 \equiv Q_{\rm P}(\theta_{\rm joint}) + Q_{\rm L}(\theta_{\rm joint}).
\end{equation}
Substituting $\theta_{\rm joint} = (F_{\rm P}+F_{\rm L})^{-1}(F_{\rm P}\theta_{\rm P} + F_{\rm L}\theta_{\rm L})$ and factoring out the displacement $\Delta\boldsymbol{\theta} = \theta_{\rm P}-\theta_{\rm L}$, the algebraic expansion simplifies exactly to
\begin{equation}
T_{\rm joint}^2
=
(\theta_{\rm P}-\theta_{\rm L})^T
\big[ F_{\rm P} (F_{\rm P}+F_{\rm L})^{-1} F_{\rm L} \big]
(\theta_{\rm P}-\theta_{\rm L}).
\end{equation}
Recognizing the tension Fisher matrix $\mathbf{F}_T = F_{\rm P} (F_{\rm P}+F_{\rm L})^{-1} F_{\rm L}$, which is algebraically equivalent to the inverse of the sum of the covariances $(F_{\rm P}^{-1} + F_{\rm L}^{-1})^{-1}$, this expression recovers the standard symmetric multi-probe Gaussian tension metric. This rigorous derivation justifies the projection of parameter displacements onto the curvature metric used throughout the main text.

\section{Numerical construction of the Fisher matrices}
\label{app:numerical_config}

The Fisher matrices used in this work are reconstructed from three independent observational probes: the Planck 2018 CMB likelihood, DESI DR2 BAO measurements, and the Pantheon+SH0ES supernova distance-ladder data set. Table~\ref{tab:numeric_config_final} summarizes the numerical configuration adopted for each probe, including the input data, Fisher construction method, parameter basis, and covariance treatment.

For the Planck analysis we use the publicly available Planck 2018 MCMC chains corresponding to the \texttt{base\_w\_plikHM\_TT\_lowl\_lowE} likelihood combination. The Fisher matrix is reconstructed from the posterior covariance of the chain samples according to
\begin{equation}
F_{\rm P} = C_{\rm P} ^{-1},
\end{equation}
where $C_{\rm P} $ denotes the marginalized covariance matrix of the parameters retained in the reduced parameter space. Early-universe parameters are marginalized over before projecting the Fisher matrix onto the late-time subspace used in the main text.

For DESI DR2 we use the BAO observables $(D_M/r_d,\,D_H/r_d)$ together with the full covariance matrix provided by the DESI collaboration.  The Fisher matrix is reconstructed from the posterior covariance of the MCMC samples used in the BAO analysis. Two treatments of the sound horizon are considered: the \textit{rdFREE} case, where $r_d$ is treated as a free parameter, and the \textit{rdFIX} case, where $r_d$ is fixed to the model-derived value.

For the Pantheon+SH0ES supernova data set the Fisher matrix is constructed numerically using central finite differences of the distance-modulus likelihood with respect to the parameters $(H_0,\Omo,M)$.  The derivatives are evaluated with the step sizes listed in Table~\ref{tab:numeric_config_final}.  The resulting Jacobian matrix $J$ enters the Fisher matrix as
\begin{equation}
F = J^{T} C^{-1} J ,
\end{equation}
with an additional Gaussian calibration prior applied to the absolute magnitude parameter $M$.

\begin{table*}[t]
\centering
\caption{Numerical configuration of the Fisher reconstructions. Only SH0ES employs finite-difference derivatives, while Planck and DESI use posterior covariance inversion.}
\label{tab:numeric_config_final}
\begin{tabular}{lccc}
\hline\hline
 & Planck 2018 (base\_w) & DESI DR2 ($D_M/r_d, D_H/r_d$) & SH0ES \\ 
\hline

Input data
& MCMC chains (\texttt{plikHM\_TT\_lowl\_lowE})
& BAO vector + covariance
& Pantheon+SH0ES .dat + .cov \\

Fisher construction
& $F_{\rm P} = C_{\rm P}^{-1}$ (posterior covariance)
& $F_{\rm D} = C_{\rm D}^{-1}$ (posterior covariance)
& $F_{\rm S} = J^T C_{\rm S}^{-1} J + \mathrm{prior}$ \\

Finite difference
& Not used
& Not used
& Central finite difference \\

Step sizes
& N/A
& N/A
& $\Delta H_0 = 10^{-3}$ \\
&  &  & $\Delta \Omo = 10^{-4}$ \\
&  &  & $\Delta M = 10^{-4}$ \\

$r_d$ treatment
& Model-derived
& rdFREE / rdFIX
& Not applicable \\

Parameter basis
& $(\Omo, \ln H_0, w)$
& rdFREE: $(\Omo, h\,r_d, w)$ \\
&  & rdFIX: $(\Omo, H_0, w)$
& $(\Omo,  H_0, M)$ \\

Burn-in
& 30\%
& 2000 steps
& Not applicable \\

Production steps
& Full chain
& 5000 MCMC steps
& Deterministic evaluation \\

Covariance
& Weighted chain covariance
& DESI full covariance
& SN covariance \\

\hline\hline
\end{tabular}
\end{table*}

\section*{Acknowledgments}
This work is supported by the Basic Science Research Program through the National Research Foundation of Korea (NRF), funded by the Ministry of Science and ICT under Grant No.~NRF-RS-2021-NR059413 and NRF-2022R1A2C1005050.

\end{document}